\documentclass[a4paper,10pt]{article}
\usepackage[utf8]{inputenc}
\usepackage[T1]{fontenc}
\usepackage{graphicx}
\usepackage{xr-hyper}
\usepackage{hyperref}       % hyperlinks
\usepackage{url}            % simple URL typesetting
\usepackage{a4wide}
\usepackage{mathtools}
\usepackage[table,dvipsnames]{xcolor}         % colors
\usepackage{mathtools}
\usepackage{authblk}

\usepackage[backend=bibtex,style=numeric-comp,sorting=none,defernumbers=true,maxcitenames=2,maxbibnames=20,sortcites=true]{biblatex}
\hypersetup{colorlinks = true, linkcolor = NavyBlue,
            urlcolor  = NavyBlue,
            citecolor = NavyBlue,
            anchorcolor = NavyBlue}

\title{Physical partisan proximity outweighs online ties in predicting US voting outcomes}

\author[1,2]{Marco Tonin}
\author[2]{Bruno Lepri}
\author[1,*]{Michele Tizzoni}

\affil[1]{Department of Sociology and Social Research, University of Trento, Trento, Italy}
\affil[2]{Mobile and Social Computing Lab (MobS), Fondazione Bruno Kessler (FBK), Trento, Italy}

\affil[*]{michele.tizzoni@unitn.it}

\begin{document}
\flushbottom
\maketitle

\begin{abstract}
Affective polarization and increasing social divisions affect social mixing and the spread of information across online and physical spaces, reinforcing social and electoral cleavages and influencing political outcomes. 
Here, using individual survey data and aggregated and de-identified co-location and online network data, we investigate the relationship between partisan exposure and vote choice in the US by comparing offline and online dimensions of partisan exposure.
By leveraging various statistical modeling approaches, we consistently find that partisan exposure in the physical space, as captured by co-location patterns, more accurately predicts electoral outcomes in US counties, outperforming online and residential exposures.
Similarly, offline ties at the individual level better predict vote choice compared to online connections. 
We also estimate county-level experienced partisan segregation and examine its relationship with individuals' demographic and socioeconomic characteristics.
Focusing on metropolitan areas, our results confirm the presence of extensive partisan segregation in the US and show that offline partisan isolation, both considering physical encounters or residential sorting, is higher than online segregation and is primarily associated with educational attainment.
Our findings emphasize the importance of physical space in understanding the relationship between social networks and political behavior, in contrast to the intense scrutiny focused on online social networks and elections.
\end{abstract}

\vspace{-1cm}
\vspace{0.5cm}

\section*{Introduction}
Affective polarization has emerged as a critical concern for contemporary democracies, dividing populations along partisan and political lines and impacting various aspects of individuals' lives \cite{iyengar2012affect,webster2017ideological,iyengar2019origins,holliday2024affective,kish2024unraveling}. 
Moreover, recent years have witnessed considerable attention to the impact of online social media and news consumption - specifically misinformation and disinformation - on opinion dynamics, political polarization, and, consequently, on political elections \cite{bessi2016social,del2016spreading,allcott2017social,badawy2018analyzing,boxell2017greater,lazer2018science,tornberg2022digital, muise2022quantifying, allcott2024effects}.
However, political polarization and partisan exposure cannot be solely explained by human behavior on social media platforms, as they are influenced by various dimensions of individuals' daily lives, including physical and online environments, the role of social networks, and existing social divisions. 

Homogeneous social networks can contribute to increased affective polarization and the maintenance of social and electoral cleavages through interactions with like-minded individuals \cite{huckfeldt2014networks,iyengar2019origins}.
In physical spaces, the debate has focused on the division of partisans at the residential level, where the grouping of like-minded people has increased at the geographical level in the United States \cite{bishop2009big, brown2021measurement,brown2023increase}. 
Specifically, individuals generally experience limited exposure to outgroup partisans, even within the same city or neighborhood \cite{brown2021measurement}. 
Furthermore, geographical sorting is significant both within and between geographical units~\cite{brown2023increase}, with larger differences observed within areas than between them~\cite{kaplan2022partisan}.
However, in terms of residential mobility, although Americans prefer politically compatible communities~\cite{gimpel2015seeking}, they do not relocate for political reasons~\cite{mummolo2017partisans}. 
Additionally, when considering individual mobility behavior, experienced partisan segregation in activity spaces varies depending on the physical space and geographic area~\cite{zhang2023human}, and online connections between geographic areas are driven by partisan homophily~\cite{mastrosavvas2024geography}.
Finally, individual survey data shows that political polarization has increased in association with offline social network homophily \cite{butters2022polarized}.
On the other hand, in online spaces and social media, affective polarization may be associated with self-selection and echo chambers \cite{del2016spreading,cinelli2021echo} or with the exposure to nonlocal interactions that may fuel political conflicts \cite{tornberg2022digital}. 
However, ideological segregation in the US is higher in face-to-face interactions than in offline and online news consumption \cite{gentzkow2011ideological}, and partisan segregation in news consumption appears to be higher in traditional media than in online social networks~\cite{gentzkow2011ideological,muise2022quantifying}, with greater heterogeneity in partisan exposure online~\cite{barbera2014social}.
Moreover, polarization has increased most significantly among social and demographic groups that are less likely to use social media \cite{boxell2017greater}, while deactivation experiments have shown no significant effects on voter turnout in US presidential elections or affective and ideological polarization \cite{allcott2024effects}. 

Although previous studies have focused mainly on partisan exposure in terms of residential locations, experienced social exposure extends beyond the place of residence and involves routines, habits, and social interactions throughout people's daily experiences~\cite{wang2018urban, moro2021mobility, zhang2023human,de2024people, nilforoshan2023human, liao2025socio}, both in physical and online spaces~\cite{dong2020segregated}.
In this context, social networks represent and shape many aspects of individual lives, with relevant impacts on access to novel information \cite{granovetter1973strength,aral2011diversity,park2018strength}, economic actions and prosperity \cite{granovetter1985economic,eagle2010network,jahani2023long}, and political participation \cite{berelson1954voting,campbell2013social,mcadam1993specifying}. 
In particular, the literature highlights how social networks, including spouses, family, close friends and acquaintances, and the social environment foster political discussion and participation \cite{campbell1960american, klofstad2009measurement, verba1995voice, lazer2010coevolution,strother2021college}, and reinforce political beliefs \cite{lazarsfeld1944people, berelson1954voting,huckfeldt1995citizens}.
Moreover, some studies have shown that individuals' vote choices are associated with the voting preferences of their political discussion networks \cite{huckfeldt1995citizens,sinclair2012social}.  
Social exposure and its influence on political behavior is represented by both context and contact.
Intergroup contact varies in depth, from casual to cooperative or selective encounters, and duration, from brief to sustained exposure \cite{nathan2023context}. 
While most studies focus on cooperative and sustained interactions and draw on contact theory \cite{allport1954nature}, research shows that proximity and casual exposure can also have significant effects on political behavior \cite{enos2014causal,enos2017space}.
Understanding partisan social exposure and the dynamics of segregation in both physical and online dimensions is crucial to studying the transmission of political choices through social networks, following social influence and voting contagion \cite{nickerson2008voting, braha2017voting}. 
However, a large-scale comparative analysis between online and physical partisan exposure in social networks and their relationship with political behavior is still lacking.

In this context, our study aims to understand the relationship between partisan exposure in online and physical spaces and voting outcomes. 
To this end, we integrate large-scale aggregated network data and individual surveys. 
Specifically, we first investigate whether partisan exposure differs across physical proximity and exposure to the same social context, online social connections, and residential sorting, both in metropolitan and non-metropolitan areas.
Second, we explore how partisan segregation at the county level relates to the demographic and socioeconomic characteristics of US counties.
Third, we evaluate the dimension that best predicts voting patterns at both the county and individual levels. 
Our study is not intended to predict the results of future presidential elections. Instead, it focuses on contrasting existing online social networks with those in the physical world to describe electoral patterns.

To compare the online and physical exposure patterns, we leverage four datasets. 
First, we consider offline social connectedness using the Colocation Maps dataset, which provides the co-location probability between two randomly selected individuals residing in different US counties \cite{iyer2023large}. Second, we analyze the Social Connectedness Index (SCI), which measures the relative probability of friendship on Facebook between two randomly selected individuals who reside in different counties \cite{bailey2018social}. 
Third, we consider the probability of being exposed to Democratic and Republican voters based on the place of residence, looking at the neighbors of the individuals, using a publicly available dataset~\cite{dati-brown}. 
We compute partisan exposure to Democrats and Republicans for each county in the contiguous United States and each dimension: offline, online, and residential. We then model the relationship with voting patterns at the county level and seek to determine which dimension best explains the observed variance of voting patterns.
Fourth, we used the 2020-2022 Social Media Study data~\cite{anes-data} provided by the American National Election Studies (ANES) to model the relationship between partisan exposure to Democrats and Republicans and the political vote in the 2020 presidential elections at the individual level. 
These datasets capture different types of contact across distinct contexts (see SI, Table S1).
Colocation Maps reflect physical co-presence, including both casual and selective, as well as brief and repeated encounters. These contacts are generally passive.
The Social Connectedness Index captures digital social ties, typically selective, sustained, and active.
Residential proximity represents the physical context in which all types of contact, casual or selective, brief or sustained, can occur. This form of contact is largely passive, shaped by where people live.
Finally, survey data provide detailed measures of interpersonal relationships, reflecting selective, sustained, and primarily active forms of social contact.

Our findings reveal the strong and significant role of physical proximity and local exposure in predicting the voting outcomes of the US counties, as well as the role of offline ties in predicting individuals' political votes in the 2020 presidential election when compared to online ties.
%Moreover, when focusing on the electoral outcomes of swing counties, where political results are less predictable from year to year, we show that physical proximity significantly outweighs all other dimensions in the prediction task. 
Moreover, we find that partisan segregation is higher in offline social networks than in online ones, with physical partisan segregation primarily associated with the county population's educational attainment and the urban-rural divide.
Overall, our work contributes to a deeper understanding of how social exposure reflects political outcomes in the US, specifically the voting patterns in presidential elections at both the county and individual levels, highlighting the effects of partisan segregation on voting polarization, with potential implications for affective polarization. 
Our results highlight the importance of gauging the relative role of different types of exposure, offline and online, when considering their effects on election results.

\section*{Results}

In this study, we compare offline and online partisan exposure and its relationship with vote choice at both the county and individual levels in the United States, and we examine partisan segregation.
We first measure partisan exposure at the county level.
Second, we analyze how partisan segregation relates to individuals' demographic and socioeconomic characteristics within metropolitan areas.
Finally, we estimate the relationship between partisan exposure and vote choice at both the county and individual levels.

\subsection*{Defining partisan exposure across different dimensions}\label{partisan-exposure}
\begin{figure}[t!]
    \centering
    \includegraphics[width=1\linewidth]{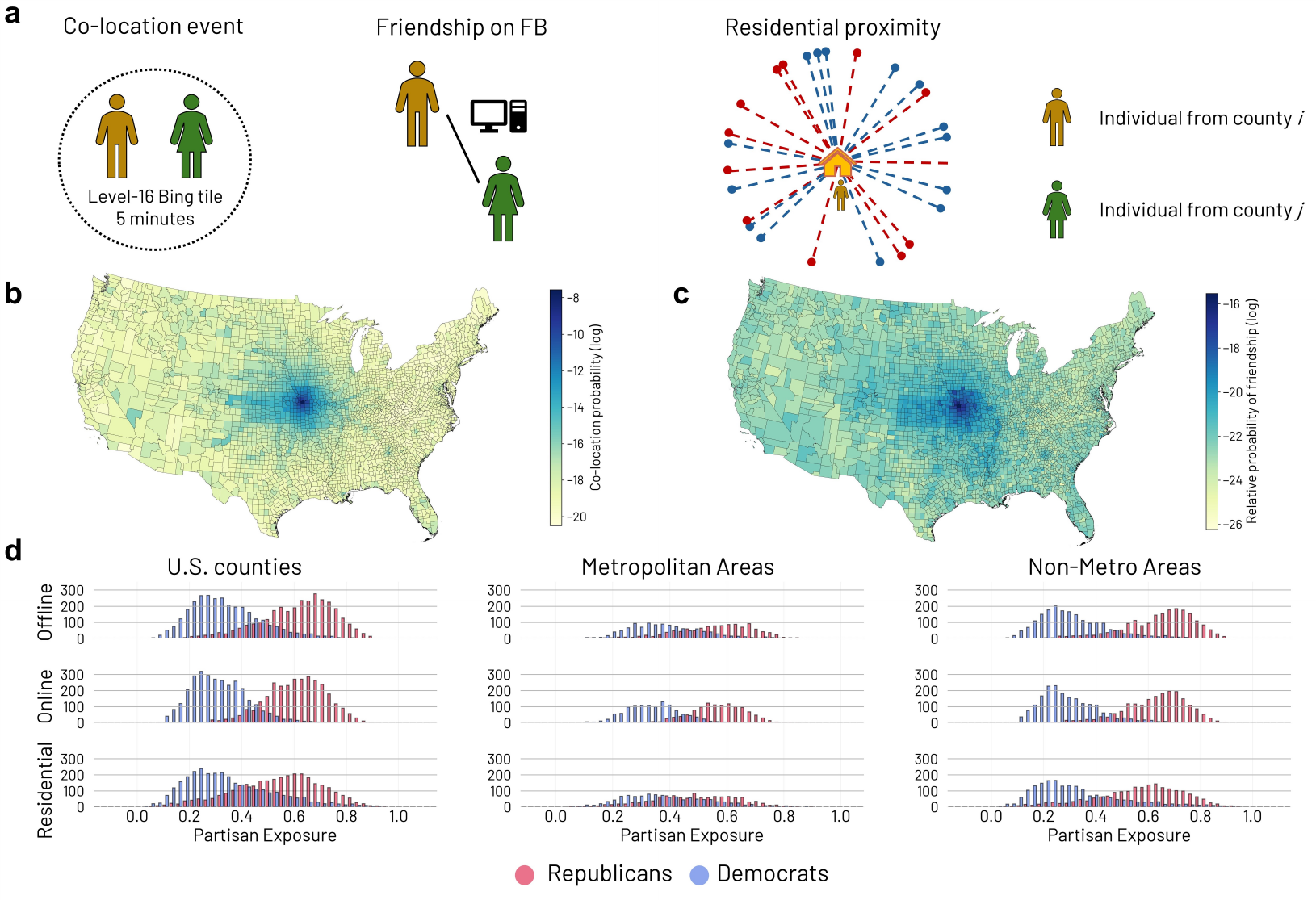}
    \caption{\textbf{The three dimensions considered to estimate partisan exposure.} (\textbf{a}) A co-location event between two randomly selected individuals from counties i and j is defined as being co-located in the same place for at least 5 minutes, while the Social Connectedness Index accounts for the number of friendships on Facebook between individuals from i and j. Residential proximity considers the nearest 1,000 individuals that registered to vote. (\textbf{b} and  \textbf{c}) Co-location probabilities and relative probabilities of friendship on Facebook, respectively, between Jackson County, MO, and all the others. (\textbf{d}) Distributions of partisan exposure by county, including metro and non-metro areas. Note: distributions are not population-weighted.}
    \label{fig1}
\end{figure}

We first compute the relative exposure to Republican and Democratic voters for each dimension. 
Physical partisan exposure is derived from the Colocation Maps, while online exposure is measured from the Social Connectedness Index (SCI). 
Specifically, the former provides the co-location probability (Fig. \ref{fig1} a) between two randomly selected individuals from US counties $i$ and $j$ \cite{iyer2023large}. 
A co-location event (Fig. \ref{fig1}a) is registered when two Facebook users are at the same location for at least 5 minutes \cite{iyer2023large}. 
In contrast, the SCI dataset provides the relative probability of friendship on Facebook between two users from $i$ and $j$ (Fig. \ref{fig1}a) \cite{bailey2018social}.
We refer to the Materials and Methods Section for further details.

Despite a high correlation between Colocation Maps and SCI raw probabilities (Pearson's $\rho = 0.83$), their corresponding networks show significant structural differences. 
Specifically, as shown in Figs. \ref{fig1}b and \ref{fig1}c, which compare the connection probabilities for Jackson County, MO, online connections are more homogeneously distributed across the country than offline co-location events. 
This is further explored in Supplementary Information, Fig. S1, which shows that US counties display a higher heterogeneity of social connections (network diversity $D(i)$) in the online space, with a median value of $0.39$ compared to $D(i) = 0.15$ offline. 
Moreover, US counties have higher online extroversion (defined as the ratio between external and internal connection probabilities) than offline extroversion, with a median value of $1.19$ compared to $0.27$.

Experienced physical and online partisan exposures, $PE$, for a county $i$ are computed as the relative exposure to Republican and Democratic voters of a county $j$ weighted by either the co-location or online connection probabilities between $i$ and $j$ (see Materials and Methods Section for more details). 
Specifically, we define the share of Republican and Democratic voters as the average share of votes for Republican and Democratic candidates in three past presidential elections (that is, 2012, 2016 and 2020), following the normal vote concept \cite{converse1966concept} to account for candidate- or election-specific influences on voter turnout and election results.
Partisan exposure ($PE$) is a measure that ranges from 0 (low exposure to Democratic or Republican voters) to 1 (high exposure to either Democratic or Republican voters).

We then calculated the residential partisan exposure for each county from the dataset provided by R. Enos \cite{dati-brown, brown2021measurement}, which provides the conditional probabilities of being exposed to Republican or Democratic voters for both Republicans and Democrats at the residential level, based on proximity to the nearest 1,000 individuals who have registered to vote. 
To the aim of our study, we compute the partisan exposure to Democratic and Republican voters for each US county as the probability that a random individual is exposed to Republicans or Democrats. Therefore, while measures of offline and online exposure are calculated from network ties between counties based on their political vote, residential exposure is derived from voter registration data. 
Further details can be found in the Materials and Methods Section. 

To investigate the differences between partisan exposure across the three dimensions, we performed t-tests and Kolmogorov-Smirnov tests between the distributions of offline, online, and residential exposures (SI, Table S2). Note that the distributions are not population-weighted. Considering all counties in the contiguous United States (Fig. \ref{fig1}d), we find significant differences ($p<.001$) in partisan exposure to Republican voters when comparing residential exposure to online or offline exposure. However, no significant difference ($p>.05$) is observed between the means of the distributions of online and offline exposures (SI, Table S3). For Democratic voters, the differences in partisan exposure in the various dimensions are also significant ($p<.001$).
Finally, we categorize counties into metropolitan and non-metro areas based on the Rural-Urban Continuum Codes (RUCC) and test the significance of the differences between partisan exposures (Fig. \ref{fig1}d; see SI, Tables S4, S5, and S6 for complete results).

\subsection*{Partisan segregation across demographic and socioeconomic factors}\label{segregation}
To investigate the levels of social segregation across political lines in the US, we estimate partisan segregation for each dimension as the net difference between exposure to Republicans and exposure to Democrats in a given county. 
The metric ranges from -1 (indicating exclusive exposure to Democrats) to 1 (indicating exclusive exposure to Republicans) and accounts for third-party voters. 
Intermediate values represent different mixing levels, with 0 indicating an equal share of exposure between the two electorates (see Materials and Methods Section for further details). 
We then investigated the relationship between partisan segregation and the demographic and socioeconomic characteristics of the county, focusing on metropolitan areas. 
To this end, we predict partisan segregation for each network dimension using Gradient Boosting (GB) regressions, based on a set of demographic and socioeconomic factors: the shares of Hispanics and Latinos, African Americans, graduated individuals, unemployment, and urban population. Furthermore, we improve the explainability of the model by computing SHAP values \cite{lundberg2017unified}, which provide importance scores that represent the impact of each variable on the predicted outcome. We refer to Materials and Methods for further details on machine learning models.

\begin{figure}[t!]
    \centering
    \includegraphics[width=1\linewidth]{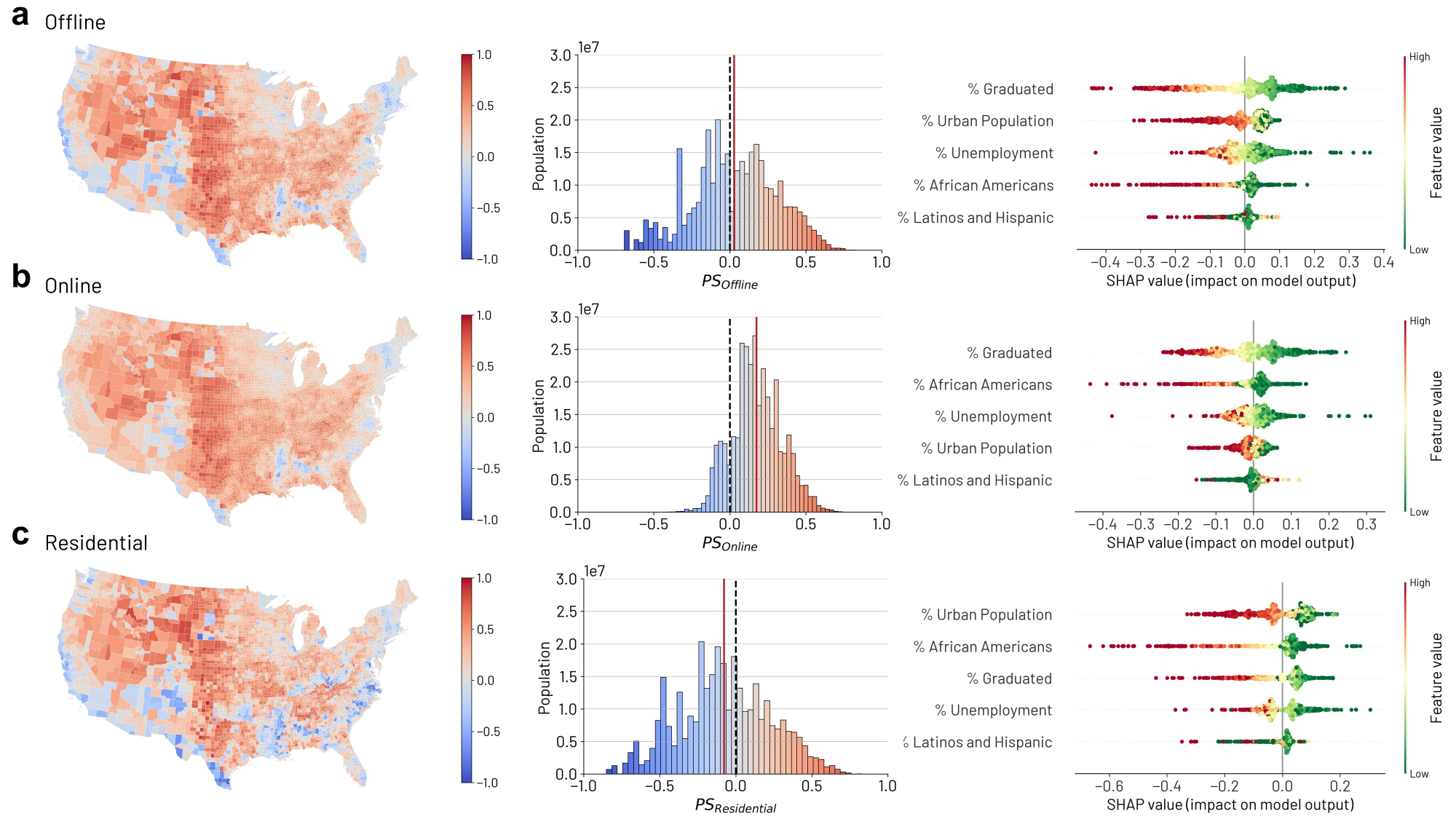}
    \caption{\textbf{Partisan segregation across demographic and socioeconomic factors for each dimension.}
    Maps and population-weighted distributions of the partisan segregation as captured by the Colocation Maps (\textbf{a}), Social Connectedness Index (\textbf{b}), and at the residential level (\textbf{c}). The red solid line represents the weighted average, while the black dashed line indicates balanced social mixing ($0$). For each dimension, the SHAP values distributions, computed from the GB regressions, highlight how the demographic and socioeconomic characteristics of the counties impact partisan segregation. Predictors are ordered by their impact on the final prediction, with points colored red for high and green for low values.}
    \label{fig5}
\end{figure}

Overall, partisan segregation in the physical world, both considering co-location maps and residential sorting, is higher than online segregation. 
As shown by the distributions of Fig. \ref{fig5}, in the offline and residential networks we observe counties characterized more frequently by $PS>0.5$ (highly Republican-segregated) or $PS<-0.5$ (highly Democrat-segregated). 
In particular, largely populated urban areas display values of $PS<-0.5$ in offline networks that are not met by online links.  

In general, GB regressions effectively explain the variance of offline segregation ($R^2=0.63$), online ($0.55$), and residential ($0.64$).
Moreover, Fig.~\ref{fig5} shows the distribution of SHAP values for each predictor related to offline, online, and residential partisan segregation in metropolitan areas. The predictors are ordered according to the impact on the prediction, with the most impactful variable at the top, and each point represents an observation. Finally, each point is colored according to its value, with red indicating high values and green indicating low values. The absolute impacts of the predictors are reported in the Supplementary Information (Table S7).

The percentage of graduated individuals is the best predictor for both partisan segregations computed from offline networks (average absolute impact equal to $0.096$) and online ($0.067$) networks. 
While counties with higher shares of graduated individuals tend to be exposed to Democrats (with higher Democratic segregation), counties with less educated individuals tend to be co-located and connected online with individuals who vote for the Republican party (Figs. \ref{fig5}a and \ref{fig5}b). 
In contrast, at the residential level, the educational level follows similar patterns but is the third predictor (mean average impact equal to $0.067$) in terms of variable importance (Fig. \ref{fig5}c). 

Moreover, we find that the shares of urban population have a greater impact on partisan segregation in offline spaces, including both co-location ($0.033$) and residential ($0.045$) exposures, compared to online connections ($0.016$). 
Specifically, metropolitan areas with urban characteristics show higher levels of Democratic segregation in offline spaces (Fig. \ref{fig5}). 

Furthermore, partisan segregation is closely related to the presence of ethnic communities. 
Specifically, the presence of a large percentage of African Americans is the second most relevant predictor of partisan segregation both online ($0.037$) and at the residential level ($0.075$) in metropolitan areas. 
Large African American communities are associated with predominant exposure to Democrats, while areas with a low African American presence tend to Republican segregation (Fig. \ref{fig5}). 
Finally, as shown in Fig. \ref{fig5}, the presence of Latinos and Hispanics is the least relevant predictor on all dimensions, namely offline ($0.022$), online ($0.017$) and at the residential level ($0.029$).
Interestingly, a larger share of Latinos and Hispanics in US counties are associated with higher levels of Democratic and Republican offline and online segregation.

\subsection*{Offline partisan exposure better predicts the voting behavior of counties}\label{prediction-counties}
We estimate the relationship between physical, online and residential partisan exposure and vote choice through several statistical models to assess their relative contribution to voting dynamics. 
As with the calculation of partisan exposure, voting patterns of US counties are defined as the average share of Democratic and Republican voters in the last past presidential elections (i.e., 2012, 2016, and 2020). The shares are not complementary because we account for third-party and null votes. 
According to the normal vote concept \cite{converse1966concept}, this method adjusts for candidate- or election-specific influences on voter turnout and election results, providing a more accurate assessment of voting patterns.
In all the models, the shares of Democrats and Republicans in the US counties are our dependent or target variables. 

We first model each dimension of partisan exposure ($PE$) and voting patterns for Democrats and Republicans separately with spatial autoregressive lag models \cite{ans88}, accounting for spatial autocorrelation across all counties in the contiguous United States.
We design distinct models due to the high correlation between the dimensions, as shown in the Supplementary Information (SI, Fig. S2).
We use k-nearest neighbor to compute spatial weights, with different values of $k$ ($5$, $7$, and $10$) for robustness.
Similarly, we evaluate the relationship between partisan exposure and voting patterns in metropolitan and non-metropolitan areas with ordinary least squares (OLS) regressions. 
Finally, we measure direct, indirect, and total effects of the independent variables for spatial models (see SI, Table S11 and Fig. S3a), as well as marginal effects for metro and non-metro areas (Fig. S3b and S3c).
See Supplementary Information for complete results (Tables S8 to S19).

\begin{figure}[t!]
    \centering
    \includegraphics[width=1\linewidth]{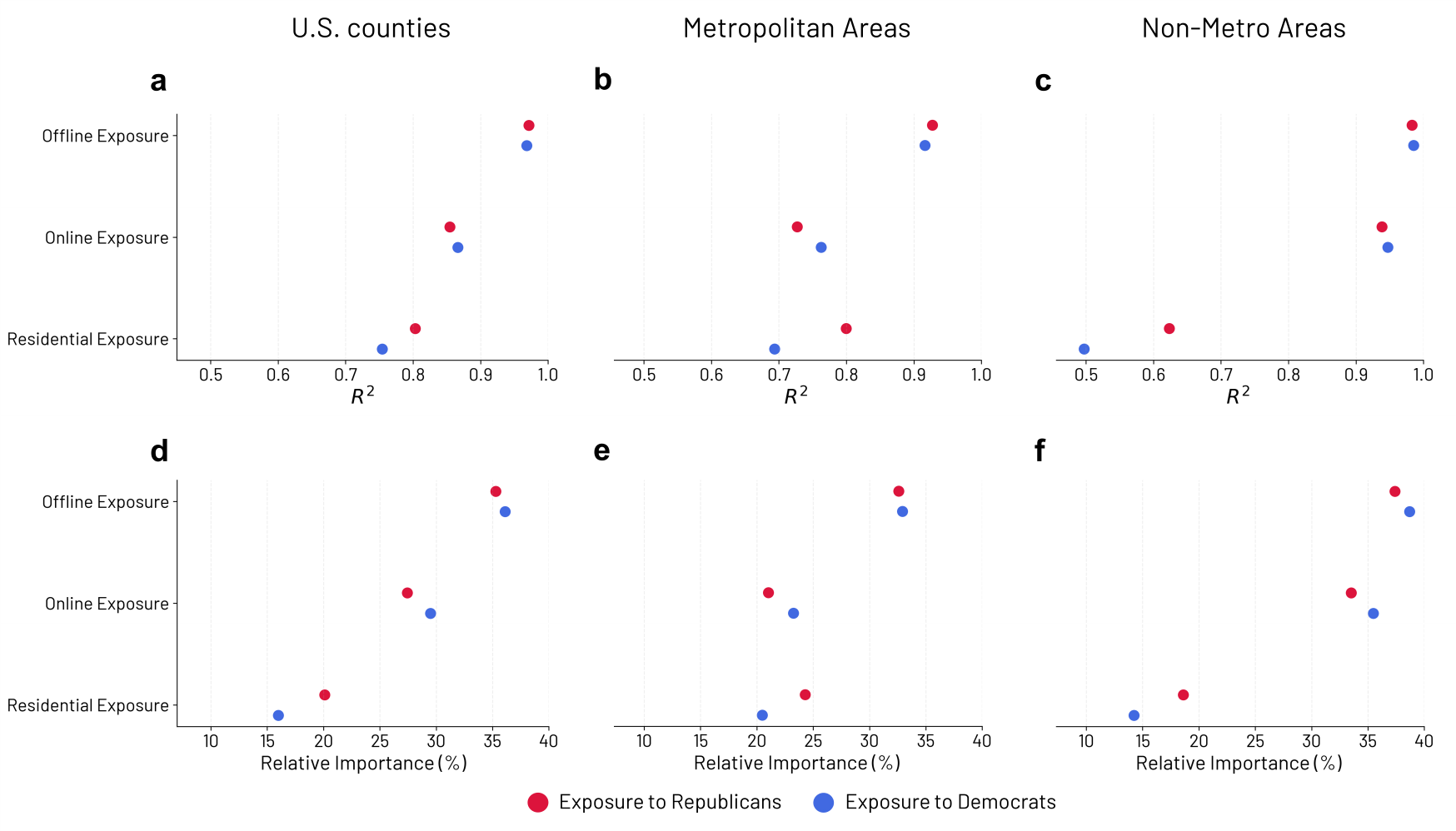}
    \caption{\textbf{Relative contribution of the three dimensions of partisan exposure on voting patterns.} 
    Physical partisan exposure outweighs online and residential exposure considering all the counties in the contiguous US in both spatial models (\textbf{a}) with $k=7$ ($R^2$) and dominance analysis (\textbf{d}) which considers demographic and socioeconomic controls. The result is consistent in both metropolitan (\textbf{b} and \textbf{e}) and non-metro areas (\textbf{c} and \textbf{f}), employing both OLS models ($R^2$) and dominance analysis.}
    \label{fig3}
\end{figure}

We estimate the relationship between physical, online and residential partisan exposure and vote choice through several statistical models to assess their relative contribution to voting dynamics. 
As with the calculation of partisan exposure, voting patterns of US counties are defined as the average share of Democratic and Republican voters in the last past presidential elections (i.e., 2012, 2016, and 2020). The shares are not complementary because we account for third-party and null votes. 
According to the normal vote concept \cite{converse1966concept}, this method adjusts for candidate- or election-specific influences on voter turnout and election results, providing a more accurate assessment of voting patterns.
In all the models, the shares of Democrats and Republicans in the US counties are our dependent or target variables. 

We first model each dimension of partisan exposure ($PE$) and voting patterns for Democrats and Republicans separately with spatial autoregressive lag models \cite{ans88}, accounting for spatial autocorrelation across all counties in the contiguous United States.
We design distinct models due to the high correlation between the dimensions, as shown in the Supplementary Information (SI, Fig. S2).
We use k-nearest neighbor to compute spatial weights, with different values of $k$ ($5$, $7$, and $10$) for robustness.
Similarly, we evaluate the relationship between partisan exposure and voting patterns in metropolitan and non-metropolitan areas with ordinary least squares (OLS) regressions. 
Finally, we measure direct, indirect, and total effects of the independent variables for spatial models (see SI, Table S11 and Fig. S3a), as well as marginal effects for metro and non-metro areas (Fig. S3b and S3c).
See Supplementary Information for complete results (Tables S8 to S19).

In Fig.~\ref{fig3}, we show the results for both spatial and OLS models, for Democrats and Republicans, further disaggregated by metropolitan and non-metropolitan counties. 
We compare the models' ability to explain the variance in the dependent variable using $R^2$, with consistent results when measuring the model's quality with the Akaike Information Criterion (AIC) and log-likelihood.
Considering all US counties and accounting for spatial autocorrelation, physical partisan exposure to both Democrats and Republicans, as captured by Colocation Maps, is the dimension that best explains the variance in the share of Democratic ($R^2=0.97$) and Republican ($R^2=0.97$) votes, respectively, compared to online ($R^2=0.87$ and $0.85$) and residential ($R^2=0.80$ and $0.75$) exposures (Fig. \ref{fig3}a). 
Moreover, when comparing online and residential partisan exposures, the former has a higher explanatory power of the variance in voting patterns than the latter for Democrats and Republicans. 
Disentangling the analysis across the urban-rural axis according to the RUCC, we find that while in non-metro areas offline and online partisan exposures have similar predictive power ($R^2=0.98$ and $0.94$ respectively for Republicans and $R^2=0.99$ and $0.95$ for Democrats), in metro areas the partisan exposure captured by Colocation Maps significantly outperforms both the other dimensions, with the $R^2$ of the model related to offline exposure equal to $0.93$ (Republicans) and $0.92$ (Democrats) compared to online ($0.73$ and $0.76$ respectively) and residential ($0.80$ and $0.69$) exposures (Figs. \ref{fig3}b and \ref{fig3}c). 
Therefore, in metropolitan areas, voting patterns are predominantly associated with partisan exposure to Democrats and Republicans in the physical space, when sharing the same social context. 

To verify the robustness of our results, we test our findings with different modeling approaches.
First, we perform a dominance analysis \cite{azen2003dominance} to compare the three dimensions, taking into account possible demographic and socioeconomic confounders, including the shares of Hispanics and Latinos, African Americans, graduated individuals, unemployment, and urban population. 
These characteristics are selected based on the literature on the determinants of political outcomes \cite{gelman_red_2010,scala2017political,ambrosius2016blue,gelman_red_2010,kahane2020determinants, kuriwaki2024geography} and by computing the variance inflation factors (VIF) to address multicollinearity issues. We refer to the Materials and Methods Section for further details about the variable selection.
From this perspective, dominance analysis allows us to deal with highly correlated variables (SI, Fig. S2) and determine the relative importance of each predictor. 
Fig. \ref{fig3}d confirms the results obtained using the spatial lag models (Fig. \ref{fig3}a), with the higher relative importance of offline exposure in explaining the variance of US voting patterns. 
Specifically, offline exposure shows greater relative importance for both Republicans ($35.29\%$) and Democrats ($36.12\%$) compared to online ($27.43\%$ and $29.5\%$, respectively) and residential ($20.11\%$ and $15.97\%$) exposures.
Moreover, the results of the OLS models (Figs. \ref{fig3}b and \ref{fig3}c) are confirmed by the dominance analysis in both metropolitan (Fig. \ref{fig3}e) and non-metropolitan areas (Fig. \ref{fig3}f). 

Second, we improve the generalizability of our findings by modeling the relationship between the three dimensions of $PE$ and the voting patterns using Random Forest and Elastic Net models with k-fold cross-validation \cite{zou2005regularization} (see Materials and Methods Section and SI, Tables S20 and S21). 
Consistently with previous results, physical exposure, as captured by Colocation Maps, has the greatest impact on prediction compared to online and residential exposures. The findings hold when considering both the variable importance in the Random Forest models and the coefficients of the Elastic Net models (see the SI, Section S7). In general, the models achieve high performance in terms of generalizability, with $R^2$ between $0.97$ and $0.98$ in the test sets for all models.

To compare our results with official statistics, we calculated partisan exposure using the 2016-2020 5-Year ACS Commuting Flows and modeled the relationship between partisan exposure and the voting patterns of the US counties. Partisan exposure, as captured by commuting flows, outperforms online and residential proximity in predicting voting patterns, but shows lower performance compared to physical exposure, as confirmed by both spatial models and dominance analysis (SI, Tables S22 to S25 and Fig. S4).

Finally, as a sensitivity analysis, we excluded the self-loops from the co-location and online connections networks. 
A self-loop represents the co-location or friendship probability between individuals in county $i$ and residents of the same county, leading to an edge between $i$ and itself. 
In the physical network, local exposure represents a substantial and fundamental aspect of an individual's exposure, with a median value of counties' extroversion equal to $0.27$. 
In contrast, while local exposure still plays a significant role in the online network, the network structure is more heterogeneous and characterized by higher external exposure (median value of extroversion equal to $1.19$). Detailed descriptions can be found in the Supplementary Information (Fig. S1).
Fig. S5 of the Supplementary Information shows that when local exposure is excluded from the calculation of $PE$, online ties outweigh physical proximity in predicting US political outcomes at the county level.
Specifically, considering all the US counties, online exposure explains $65\%$ and $67\%$ ($R^2$) of the variance of voting patterns for Republicans and Democrats, respectively, compared to $61\%$ and $60\%$ explained by the physical proximity (Fig. S5a). 
However, predictive performance is lower when local exposure is not considered, particularly in the offline network, characterized by low heterogeneity and extroversion. 
Hence, local exposure appears to be a fundamental dimension in predicting voting outcomes.

\subsection*{Offline partisan exposure better predicts individuals' voting behavior}\label{individuals}

\begin{figure}[t!]
    \centering
    \includegraphics[width=0.9\linewidth]{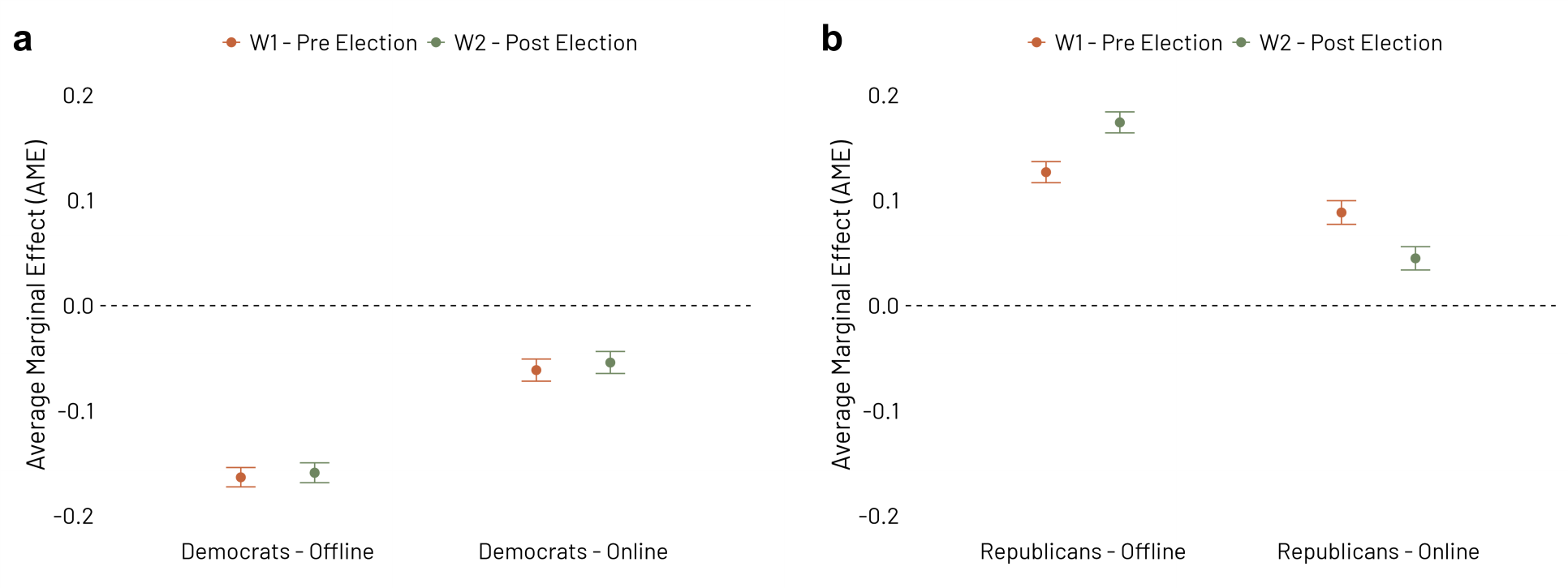}
    \caption{\textbf{Average marginal effects of partisan exposure (online and offline) on vote choice.}
    Offline partisan exposure has a stronger average marginal effect on vote choice than online partisan exposure, both in terms of exposure to Democrats (\textbf{a}) and Republicans (\textbf{b}). Logit models control for respondents' age, ethnicity, educational attainment, and place of residence (metro or non-metro area). The dependent variable is binary, with 0 for voting Democrat and 1 for voting Republican.}
    \label{fig-survey}
\end{figure}

We combine large-scale analysis with individual survey data to better understand the relationship between partisan exposure in physical and digital spaces and vote choice. 
To this end, we leverage the 2020-2022 Social Media Study dataset \cite{anes-data} provided by the American National Election Studies (ANES), which consists of an online survey panel conducted in three waves (pre- and post-election 2020, and the 2022 midterm elections). 
As explained in the Materials and Methods Section, our focus is on the first two waves, including all respondents who declared their vote choice in the 2020 presidential election for either the Democratic or Republican party, had a valid vote, and responded to the following questions: "How many of your friends and family are Democrats?", "How many of your friends and family are Republicans?", "How many of your Facebook friends are Democrats?", and "How many of your Facebook friends are Republicans?". The final sample consists of 2420 respondents. Descriptive information about the variables can be found in the Supplementary Information (Table S26).

As described in the Materials and Methods section, we performed four logit models to distinguish between exposure to Democrats and Republicans and to account for each wave (pre- and post-election). The dependent variable is vote choice, encoded as a binary variable with 0 for the Democratic party and 1 for the Republican party. We control for the age, ethnicity, educational attainment and place of residence of the respondents (metro or non-metro area), and compute and apply post-stratification weights based on gender, age, education and ethnicity using the 5-year estimates 2020-2024 of the American Community Survey (ACS). Detailed results can be found in the Supplementary Information (S27).

Fig. \ref{fig-survey} shows the Average Marginal Effects (AME) of offline and online partisan exposure on vote choice across the two waves. As shown, offline exposure to Democrats has a stronger effect on voting for the Democratic party than online exposure (\ref{fig-survey}a), while offline exposure to Republicans more significantly impacts voting for the Republican party (\ref{fig-survey}b).
Specifically, in the first wave, the average marginal effect of the exposure to Democrats is $-0.163$ offline compared to $-0.061$ online, while in the second wave it is $-0.159$ compared to $-0.054$. Similarly, related to the exposure to Republicans, the average marginal effect is $0.127$ offline compared to $0.089$ online in the first wave, and $0.174$ compared to $0.045$ in the second wave.

\section*{Discussion}

In this study, we compared partisan exposure across physical and online spaces, evaluated the association between partisan segregation in online and physical spaces and the underlying counties' demographic and socioeconomic characteristics, and investigated the relationship between partisan exposure and vote choice at both the individual and county levels. 
Our findings highlight the dominant role of experienced physical proximity and local exposure in predicting vote choice in the United States at both the individual and county levels, outperforming the influence of online ties and residential proximity. 
The result was consistent across both metropolitan and non-metro areas.
%, as well as in swing and non-swing counties. 
Moreover, we found that partisan segregation is higher in offline social networks than online ones, and such a difference is primarily driven by individuals' educational attainment. 
The urban-rural divide significantly shapes offline and residential partisan segregation, with Republican voter segregation more prevalent in metro areas with large rural populations. 
Thus, physical partisan exposure, which is the best predictor of voting outcomes, is also strongly associated with educational attainment and urbanization levels.
From this perspective, our findings contribute to a better understanding of the relationship between social networks, physical and online spaces, and political behavior, highlighting the effects of partisan segregation on voting polarization, with potential implications for affective polarization \cite{holliday2024affective}.

Despite significant attention to the impact of new media technologies, particularly online social media, on affective polarization and political elections \cite{bessi2016social,del2016spreading,allcott2017social,badawy2018analyzing,boxell2017greater,lazer2018science,tornberg2022digital, muise2022quantifying, allcott2024effects}, either through mechanisms of echo chambers and selective exposure \cite{del2016spreading, cinelli2021echo} or through interactions outside of one's local bubbles \cite{tornberg2022digital}, our findings highlight the greater relevance of physical space and offline ties in predicting political behavior. 
This is demonstrated through both large-scale (i.e., county-level) and individual survey analyses, enhancing our understanding of the relationship between partisan exposure and political vote. 
From this perspective, offline social networks, local exposure and physical proximity, characterized by low heterogeneity of social exposure and low extroversion, better reflect vote choice at the individual and county levels compared to online ties.
Similar to prior research \cite{enos2014causal,enos2017space}, our results suggest that even brief, passive, and casual forms of exposure, which can be captured by co-location events, can influence or be associated with political behavior. 
Therefore, beyond the form of exposure, context matters, with offline contexts outweighing online exposure.
In line with the literature that examines the association between social networks and vote choice \cite{huckfeldt2004political,sinclair2012social}, these findings confirm the link between homogeneous social networks and political vote, particularly in physical space, where Americans experience greater partisan segregation. This underscores the persistent influence of offline social ties compared to online ones, despite increasing attention to social media platforms and their potential impact on vote choice and affective polarization.

In general, partisan exposure and segregation vary between physical and online spaces, as well as between urban and rural areas. 
In line with the literature, our findings highlight that up to 10\% of the American population experiences a low exposure to outgroup partisans, not only in the residential sphere \cite{brown2021measurement}, but also in physical social encounters. 
Our analysis shows that experienced segregation is even greater than online partisan segregation, consistent with previous studies \cite{gentzkow2011ideological, muise2022quantifying, barbera2014social}.
Both forms of experienced segregation are associated with the educational attainment of individuals, a known driver of socioeconomic segregation \cite{browning2017socioeconomic}. 
Furthermore, the tendency of Republican voter segregation in areas with large rural populations reveals differences between urban and rural areas related not only in terms of political party support \cite{gimpel2020urban}, but also in partisan exposure \cite{brown2021measurement}. In contrast, consistent with the literature, our findings show greater Democratic segregation in high-density urban areas \cite{brown2021measurement}.

Our study comes with some limitations. 
First, we compared online and physical networks using aggregated observational data from social media platforms. 
Despite the two-step reweighting applied to Colocation Maps \cite{iyer2023large}, the Facebook data might underrepresent certain categories of people (e.g., young and old individuals) and the population of rural areas \cite{rama2020facebook}. Moreover, these aggregated networks are considered at a rather coarse geographical level, the US county, whose population sizes vary widely, ranging from $83$ individuals (Loving County, TX) to over $10$ million inhabitants. 
Furthermore, we use the SCI as a proxy of online connections, but people interact through multiple social media platforms and have access to novel information from different sources. However, although some studies have attempted to spatially map online interactions \cite{dong2020segregated, morales2019segregation, tizzoni2015scaling}, achieving comprehensive coverage throughout the US remains challenging.
Moreover, we acknowledge that co-location does not imply a face-to-face interaction between individuals. However, co-presence temporal networks have been found to have structural and statistical features comparable to face-to-face interaction temporal networks \cite{genois2018can}.
Additionally, individuals' partisan exposure, as measured by the ANES Social Media Study both before and after the 2020 US presidential election, reflects self-perceived partisan exposure and may be subject to self-report biases, such as social desirability or recall inaccuracy.
Finally, as mentioned, these datasets capture different types of contact, varying in duration, selectivity, and the extent to which they are active or passive. This is also reflected in the mismatch between county- and individual-level analyses, with the latter capturing more active forms of exposure, typically driven by friendship or kinship ties. Future work should aim at measuring the partisan balance of individual's social networks, online and offline, based on one comprehensive data source.

In conclusion, our study underscores the centrality of physical space in understanding human and political behavior, with potential implications for affective and voting polarization driven by partisan segregation. Despite challenges related to data limitations and privacy concerns, future studies on political polarization, social networks, and partisan segregation, and their impact on political outcomes, should not overlook the importance of real-world interactions in physical space.

\section*{Materials and methods}

\subsection*{Data sources}

\paragraph{Online network}
We use the Social Connectedness Index (SCI) \cite{bailey2018social} dataset provided by Meta through its Data for Good program as a proxy of online connections. 
The SCI measures the social connectedness between administrative areas by computing the probability at which two random individuals, living respectively in administrative areas $i$ and $j$, are friends on Facebook, as of April 2016 \cite{bailey2018social}. 
The SCI is available at the county and zip code levels. We analyze online connections at the county level to enable comparison with the Colocation Maps.
Specifically, to be able to compare the online probabilities with the offline ones, we use the original version of the dataset adopted in Bailey et al. \cite{bailey2018social}, which includes the relative probabilities of friendship between counties computed as follows: 
\begin{equation}
rel\_prob\_friend = C \frac{f_{ij}}{pop_i * pop_j},
\end{equation}
where $f_{ij}$ is the number of Facebook friendships between administrative regions $i$ and $j$ and $pop_i$ and $pop_j$ are the population size of $i$ and $j$. 
$C$ is a scaling factor, equal to $10^{12}$, which we drop to obtain the raw probabilities.

\paragraph{Offline network}
We use the Colocation Maps \cite{iyer2023large} provided by Meta through its Data for Good program to measure offline exposure. 
Similarly to the SCI, the dataset provides the weekly co-location probability between two randomly selected individuals from two administrative areas $i$ and $j$. 
A co-location event (Fig. \ref{fig1}a) is registered when two individuals who enabled location-based services on the Facebook smartphone application are located in the same place (i.e., a level-16 Bing-tile, approximately 600 by 600 meters at the Equator) at the same time for at least 5 minutes. 
The co-location probability between geographic regions $i$ and $j$ is then determined by the number of co-location events between individuals from $i$ and $j$ weighted by their population sizes. As described by Iyer et al. \cite{iyer2023large}, the authors applied a two-step reweighting to enhance representativeness. Specifically, the first step aligns the Facebook user data with the general population, and the second step reweights the Colocation Maps population to the Facebook user base.
The dataset has been designed and used mainly for epidemiological purposes \cite{fritz2022interplay,delussu2023limits}. For the aim of our study, we collected weekly co-location probabilities between US counties from August 2022 to January 2024, for a total of $76$ weeks/observations. 
We then use the average co-location value over the full period to mitigate seasonality effects and data sparsity and reduce computational errors in data measurement, ensuring a more realistic measure of physical exposure between counties.

\paragraph{Residential partisan exposure}
We compare online and offline networks with a measure of residential proximity, based on the residential locations of individuals. 
To this purpose, we use the dataset provided by R. Enos \cite{dati-brown} supporting the recent study by Brown et al. \cite{brown2021measurement}, which includes the conditional probability of exposure to Democratic and Republican voters, conditioned on an individual's political affiliation (either Democrat or Republican). 
The probabilities are computed based on the nearest 1,000 individuals, but the results remain consistent even considering a larger sample size of 50,000 individuals. The dataset is derived from voter registration records, which require individuals in the majority of US states to declare their party affiliation and it provides information at different geographical levels. 
To have a measure of partisan exposure at the residential level, we compute the probabilities of exposure to Democrats and Republicans for a random individual for each US county. Specifically, from the available raw data, we compute the partisan exposure to Republicans for a county $i$ as follows:
\begin{equation}
P (R) = \frac {P (R|D) P(D)} {P (D|R)},
\end{equation}
where $P (R|D)$ is the conditional probability of exposure to Republicans being a Democrat and $P (D|R)$ is the conditional probability of exposure to Democrats being a Republican. Similarly, we compute the partisan exposure to Democrats.

\paragraph{ANES 2020-2022 Social Media Study}
We analyze individual-level exposure to Democrats and Republicans and their relationship with political voting in the 2020 presidential election using the 2020-2022 Social Media Study provided by the American National Election Studies (ANES) \cite{anes-data}. 
The study is based on an online survey panel of 3 waves (that is, before and after the 2020 elections, and the 2022 midterm elections). We focus on the first two waves, which include 5,750 and 5,277 respondents, respectively. 
From this sample, we consider only those respondents who declared their political preference in the second wave ($w2presvtwho$), have valid voting records ($vote20\_match$), and answered all the questions of interest. The final sample consists of 2420 respondents.
Specifically, we focus on the following questions: ``How many of your friends and family are Democrats?", ``How many of your friends and family are Republicans?",``How many of your Facebook friends are Democrats?" and ``How many of your Facebook friends are Republicans?" Each of these questions is measured on a scale from 1 to 5, where 1 indicates ``None or almost none", and 5 indicates "All or nearly all".
We then compute new post-stratification weights based on gender, age, education, and ethnicity using the 2020-2024 5-year estimates of the American Community Survey (ACS). 

\paragraph{Voting patterns in US counties}
We define the voting patterns of counties in the contiguous United States following the normal vote concept \cite{converse1966concept} to account for candidate- and election-specific influences on voting outcomes. 
Specifically, the share of Republican and Democratic voters for each county is computed as the average of three presidential elections (i.e., 2012, 2016, and 2020).
We obtained data on presidential election results from the work by Algara and Amlani \cite{DVN/DGUMFI_2021}.

\paragraph{Demographic and socioeconomic indicators}
To evaluate the relationship between partisan segregation and county characteristics, we obtain data from 5-year estimates of the American Community Survey (ACS) for 2017-2021 provided by the US Census Bureau \cite{census}. 
Specifically, we used data related to the share of Hispanics and Latinos, African Americans, unemployed individuals, and urban population.
In addition, we use information about the share of graduated individuals for each county provided by Chetty et al. \cite{chetty2018opportunity} through the Opportunity Insights repository \cite{opportunity-insights}. 
We select this information from a broader set of counties' characteristics by computing Variance Inflation Factors (VIFs) to address multicollinearity issues and drawing on the literature on the determinants of county outcomes in US presidential elections. 
Specifically, ethnicity is one of the strongest predictors of political outcomes, and African Americans tend to vote for the Democratic party \cite{gelman_red_2010, kuriwaki2024geography}. Furthermore, educational attainment, which is highly correlated with income (addressed through VIF), is associated with Democratic voting tendencies \cite{gelman_red_2010,ambrosius2016blue}. Finally, unemployment is one of the economic indicators that affect political outcomes \cite{kahane2020determinants}, as well as the characteristics and divide of urban-rural regions \cite{scala2017political, gimpel2020urban}.

\paragraph{Metropolitan and Non-Metropolitan areas}
We define metropolitan and non-metropolitan areas based on the Rural-Urban Continuum Codes (RUCC) 2013 \cite{rucc}. Specifically, metropolitan areas include the codes 1 to 3 (RUCC 1--3), while non-metropolitan areas include codes 4 to 9 (RUCC 4--9). 

\subsection*{Methods}

\paragraph{Partisan exposure}
We define the partisan exposure of a county $i$ as the relative exposure to people with a certain voting behavior (i.e., Republican or Democratic voters), weighted by either the co-location probability between counties $i$ and $j$ (i.e., Colocation Maps) or the relative probability of friendship on Facebook between $i$ and $j$ (i.e., the Social Connectedness Index). Formally, we define the partisan exposure as:

\begin{equation}
PE_{p}(i) = \sum^{N}_{j}\ v_{p} (j)\ \frac {p_{ij}} {\sum^{N}_{k} p_{ik}}\,,
\end{equation}
where $PE_{p} (i)$ is the partisan exposure to either Republicans or Democrats for a county $i$, $N$ is the number of US counties, $v_{p} (j)$ is the percentage of Republican or Democratic voters in the county $j$, and $p_{ij}$ is either the co-location probability between counties $i$ and $j$, or the relative probability of friendship on Facebook between $i$ and $j$. 
Partisan exposure ranges from 0 (low exposure to either Democrats or Republicans) to 1 (high exposure to either Democrats or Republicans). 
The sum of partisan exposures to Republicans and Democrats is less than $1$ because we account for null and third-party votes.

\paragraph{Residential and experienced partisan segregation}
Residential and experienced partisan segregation are computed from the residential and online or offline partisan exposures, respectively. 
Specifically, partisan segregation is computed as the difference between exposure to Republicans and Democrats. 
The index ranges from -1 (indicating exclusive exposure to Democrats) to 1 (indicating exclusive exposure to Republicans) and accounts for third-party voters (since the sum of $PE_{rep} (i)$ and $PE_{dem} (i)$ may be less than $1$). Intermediate values indicate varying levels of partisan mixing. 
The index is computed as follows:
\begin{equation}
PS (i) = PE_{rep} (i) - PE_{dem} (i),
\end{equation}
where $PS (i)$ is the index of partisan segregation for a county $i$, $PE_{rep} (i)$ is the partisan exposure to Republicans and $PE_{dem} (i)$ is the partisan exposure to Democrats.

\paragraph{Spatial Autoregressive Lag Models and OLS}
To explore which dimension of partisan exposure better predicts voting patterns in US counties, we separately model physical, online, and residential exposures using spatial regressions, both for Republicans and Democrats. 
Due to the high correlation between these dimensions (see SI, S2A), combining them in the same model could impact interpretation and potentially lead to misleading conclusions.
Specifically, we use spatial lag models \cite{ans88}, which are autoregressive models that account for spatial autocorrelation among observations. Specifically, the models are defined as:
\begin{equation}
v_{p} (i) = \alpha + \rho Wy + \beta PE_{p,d} (i)+ \epsilon,
\end{equation}
where $v$ is the share of either Republican or Democratic votes, $p$ is either Democrats or Republicans, $\rho Wy$ is the autoregressive coefficient, and $PE$ is the partisan exposure to either Democrats or Republicans, $p$. The model is fitted for each dimension $d$, that is, offline, online and residential.
To enhance the robustness of the method, we compute the spatial weights of k-nearest neighbors $W$ using different values of $k$, specifically $5$, $7$, and $10$. 
This approach is preferred because contiguity-based methods would exclude non-contiguous counties from the analysis.
When comparing metropolitan and non-metropolitan areas based on the Rural-Urban Continuum Codes (RUCC), we use ordinary least squares (OLS) regressions. 
In both types of models, we determine the dimension that better predicts voting patterns based on its ability to explain the variance of the dependent variable (i.e., $R^2$).

\paragraph{Dominance analysis}
While we treat the three dimensions separately in the regressions, due to their high collinearity, we use the dominance analysis \cite{azen2003dominance} to consider them together and evaluate their relative importance.
This method allows us to evaluate the impact of predictors by measuring the contribution of each predictor through designing a series of linear regressions with all possible combinations of predictors to systematically assess and compare the individual and combined predictive power of each variable. 
Specifically, we perform a dominance analysis for both the exposure to Republicans and Democrats, including physical, online, and residential exposure in the models along with the demographic and socioeconomic characteristics previously described. 
The method provides the relative importance of each predictor as a percentage (Fig. \ref{fig3}). The full model is defined as:
\begin{equation}
v_{p} (i) = \alpha + \beta_{1} PE_{p,offline} (i) + \beta_{2} PE_{p,online} (i) + \beta_{3} PE_{p,residential} (i) + \beta_{n} controls (i) + \epsilon,
\end{equation}
where we also account for the demographic and socioeconomic characteristics of the counties.

\paragraph{Logit regressions on survey data}
To understand the relationship between partisan exposure and vote choice, we model four logit regressions, differentiating between exposure to Democrats and Republicans for each wave (pre- and post-election). This approach accounts for the complementary nature and negative correlation between questions like "How many of your friends and family are Democrats?" and "How many of your friends and family are Republicans?" The logistic regression models are specified as follows:
\begin{equation}
v(i) = \alpha + \beta_{1} PE_{offline} (i) + \beta_{2} PE_{online} (i) + \beta_{n} controls (i) + \epsilon,
\end{equation}
where $v(i)$ is the respondents' vote at the 2020 presidential elections, $PE_{offline}$ represents offline partisan exposure based on friends and family voting behavior, and $PE_{online}$ represents online partisan exposure based on Facebook friends' vote choice. We include controls for respondents' age, ethnicity, educational attainment, and place of residence (metro or non-metro area), and apply post-stratification weights based on gender, age, education, and ethnicity using the 2020-2024 5-year estimates of the American Community Survey (ACS). Ethnicity and educational attainment are treated as dummy variables, differentiating between the White, Black, and Hispanic categories and between individuals with and without a degree.

\paragraph{Predictive modeling and model explainability}
To enhance the robustness and generalizability of the result related to the relationship between partisan exposure and voting patterns, we model each dimension using Random Forest and Elastic Net \cite{zou2005regularization} models with k-fold cross-validation ($k=5$). This allows us to evaluate variable importance using tree models and beta coefficients of the model using a regularization technique.

Second, we predict partisan segregation in metropolitan areas (RUCC 1--3) for each dimension using demographic and socioeconomic characteristics of US counties. 
To this aim, we model Gradient Boosting regressions and compute SHAP \cite{lundberg2017unified} values to interpret the predictions. Specifically, drawing on game theory, the SHAP framework computes importance scores for each predictor, representing the impact of each variable on the final prediction. 
Before training the model, we split the dataset into training and test sets using a 70/30 ratio.

\section*{Acknowledgments}
We sincerely thank Michael Bailey for his assistance in accessing the original version of the Social Connectedness Index and Ryan Enos for his support with the dataset on residential partisan exposure. We also thank Alberto Acerbi and Mario Quaranta for their comments on an early version of the manuscript.

\section*{Funding}
M.To. acknowledges the support of the NRRP MUR program funded by the NextGenerationEU.
B.L. acknowledges the support of the PNRR ICSC National Research Centre for High Performance Computing, Big Data and Quantum Computing (CN00000013), under the NRRP MUR program funded by the NextGenerationEU.
The work of B.L. was partially supported by the following projects: Horizon Europe Programme, grant \#101120237-ELIAS and grant \#101120763-TANGO. Funded by the European Union. Views and opinions expressed are however those of the author(s) only and do not necessarily reflect those of the European Union or the European Health and Digital Executive Agency (HaDEA). Neither the European Union nor the granting authority can be held responsible for them.

\section*{Author contributions statement}
M.To. collected and analyzed the data, performed the research, and wrote the first version of the manuscript.
M.To., B.L., M.Ti. designed the study, interpreted the results, provided critical feedback, helped shape the manuscript, and revised it.
All authors approved the final version of the manuscript.

\section*{Data availability}
Colocation Maps are available through the Meta Data for Good program by signing a data sharing agreement (\url{https://dataforgood.facebook.com/}). The Social Connectedness Index (SCI) is publicly available through the Humanitarian Data Exchange portal (\url{https://data.humdata.org/dataset/social-connectedness-index}). Residential partisan exposure is available through Harvard Dataverse (\url{https://doi.org/10.7910/DVN/A40X5L}). The ANES 2020-2022 Social Media Survey dataset is available at \url{https://electionstudies.org/data-center/2020-2022-social-media-study/}. The code for reproducing the study findings is available at \url{https://github.com/tonmarco}.

\section*{Competing interests}
The authors declare no competing interests.

\printbibliography
% \bibliographystyle{unsrt}
% \bibliography{sample}
\end{document}

\maketitle
\tableofcontents

\newpage

\section{Data sources and type of contacts or encounters}

We compare partisan exposure across physical and digital spaces using datasets that capture different types of contact or encounters. Table \ref{tab:data_sources_exposure} details the data sources, units of analysis, and what each represents.
As discussed in the manuscript, although most of the literature has focused on cooperative and sustained contact, drawing on contact theory \cite{allport1954nature}, even brief and casual encounters can influence political behavior \cite{enos2014causal, enos2017space, nathan2023context}.

\setlength{\tabcolsep}{6pt}
\renewcommand{\arraystretch}{1}
\begin{table}[h]
    \centering
    \small  
    \caption{Data sources and types of contacts and encounters}
    \resizebox{\textwidth}{!}{%
    \begin{tabular}{@{}llp{8cm}@{}}  
        \hline
        Data Source & Unit of Analysis & Type of Encounter \\
        \hline
        Colocation Maps & Probability & 
        {Co-location for at least 5 minutes. It may be brief or sustained, casual or cooperative/selective. Generally passive.} \\
        
        Social Connectedness Index & Probability & 
        {Digital social ties on Facebook, typically selective, sustained, and active.}  \\
        
        Residential exposure & Probability & 
        {Exposure to neighbors. All types of contact, casual or selective, brief or sustained, can occur. Largely passive.} \\
        
        ANES survey & Likert scale & 
        {Offline and online interpersonal relationships, reflecting selective, sustained and primarily active forms of social contact.} \\
        \hline
    \end{tabular}
     }
    \label{tab:data_sources_exposure}
\end{table}

\clearpage
\section{Diversity and extroversion in online and offline networks}

The heterogeneity of social connectedness is defined as a function of Shannon's entropy. The measure provides the degree of diversity of counties' social connections, taking values from 0 (low diversity) to 1 (high diversity). 

\begin{equation}
D(i) = - \frac {\sum_{j=1}^{k} p_{ij} \log{(p_{ij})}} {\log{k}},
\end{equation}
where $k$ is the number of counties, and $p_{ij} = \frac{V_{ij}}{\sum_{j=1}^{k}V_{ij}}$ where $V_{ij}$ is either the co-location probability between $i$ and $j$, or the relative probability of friendship on Facebook between $i$ and $j$.

The extent of external exposure is defined with a measure of extroversion by computing the ratio between external and internal probabilities for each county, both for the co-location and friendship networks. The measure is scaled between 0 (low extroversion) and 1 (high extroversion). 

\begin{equation}
E(i) = \frac{p_{int}}{p_{ext}},
\end{equation}
where $p_{int}$ is either the co-location or friendship probability between a county $i$ and itself, and $p_{ext}$ is the sum of either the co-location or friendship probabilities of the county $i$ with all the other counties.

\begin{figure}[htp]
    \centering
    \includegraphics[width=1\linewidth]{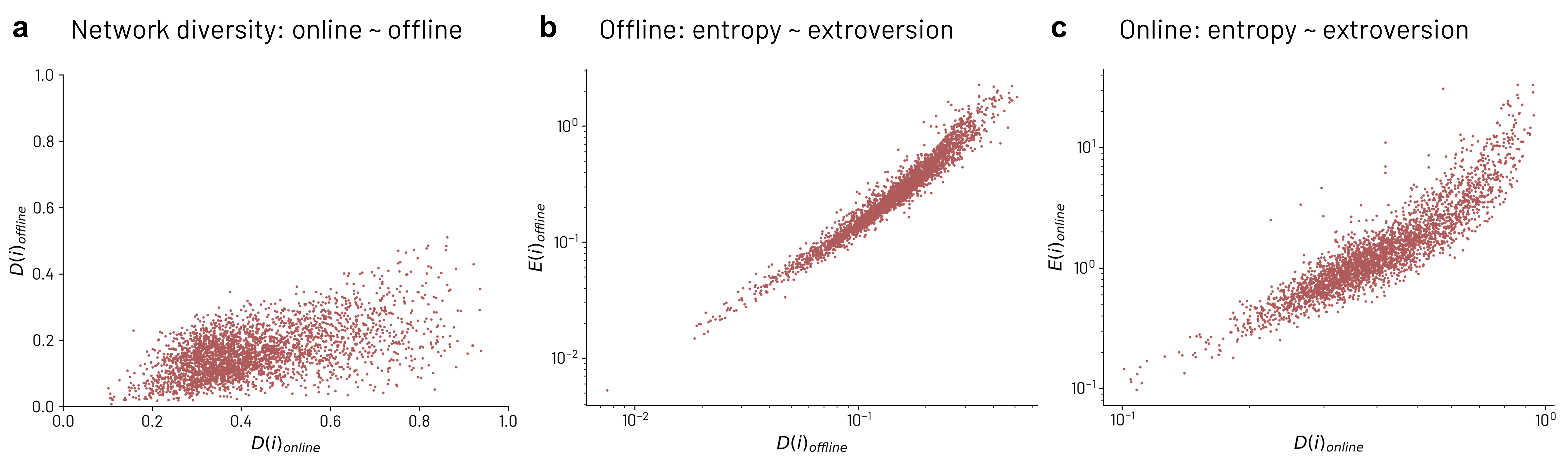}
    \caption{\textbf{Heterogeneity and extroversion in the online and offline networks.} 
    (\textbf{a}) Relationship between online and offline network diversity. (\textbf{b}) Relationship between offline network diversity and extroversion. (\textbf{c}) Relationship between online network diversity and extroversion.}
    \label{SI-fig-entropy}
\end{figure}

As shown in Fig. \ref{SI-fig-entropy}, the links in the online network are more heterogeneous compared to the offline network (See Fig. \ref{SI-fig-entropy}a). Additionally, network diversity is strongly correlated with the degree of extroversion in social connections for both online and offline networks (See Fig. \ref{SI-fig-entropy}b and \ref{SI-fig-entropy}c). 

\clearpage
\section{Partisan exposure across dimensions}

\subsection{Correlation between partisan exposure across dimensions}
\begin{figure}[htp]
    \centering
    \includegraphics[width=0.9\linewidth]{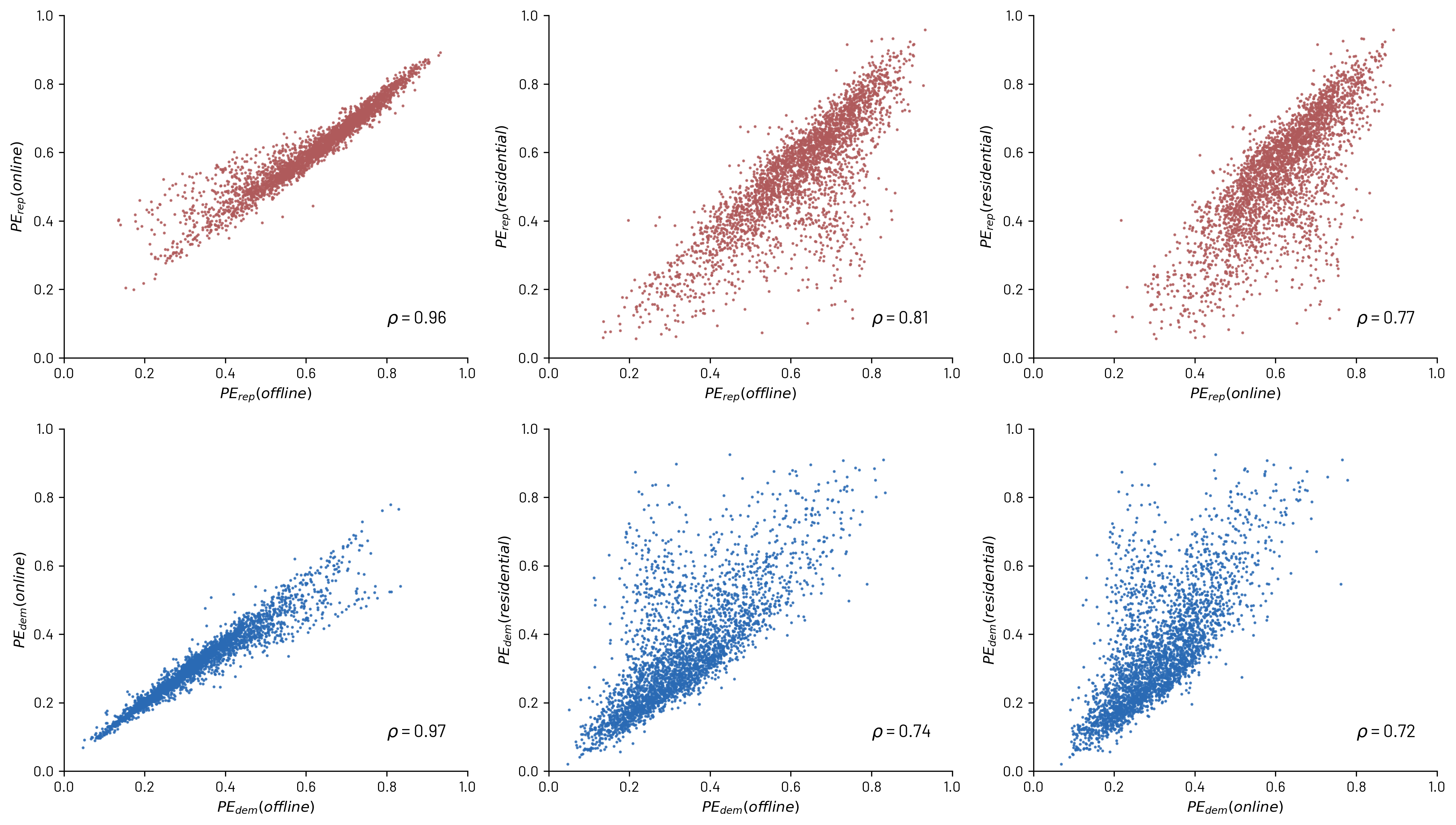}
    \caption{\textbf{Correlation between dimensions.}}
    \label{SI-fig-correlation}
\end{figure}

As mentioned in the manuscript, partisan exposure across physical and online spaces are highly correlated. Specifically, as shown in Fig. \ref{SI-fig-correlation}, the correlation between online and offline partisan exposure is equal to $0.96$ for Republicans and $0.97$ for Democrats. The correlation between offline and residential is $0.81$ and $0.74$ respectively. Finally, the correlation between online and residential partisan exposure is $0.77$ and $0.72$ respectively.

\clearpage
\subsection{Differences in partisan exposure across dimensions}
We assess the differences in partisan exposure across physical and online spaces, as well as between metropolitan and non-metropolitan areas. To this aim, we perform t-tests and Kolmogorov-Smirnov tests to compare the different dimensions and Welch's t-tests and Kolmogorov-Smirnov tests to evaluate the differences between metropolitan and non-metropolitan areas.

\setlength{\tabcolsep}{40pt}
\renewcommand{\arraystretch}{1} 
\begin{table}[h]
    \centering
    \caption{Descriptive statistics of partisan exposure}
    \resizebox{\textwidth}{!}{%
    \begin{tabular}{@{}lcccc@{}}
        \hline
        Variable & Mean & SD & Min & Max \\
        \hline
        $PE_{rep}$(offline)      & 0.620 & 0.138 & 0.134 & 0.933 \\
        $PE_{rep}$(online)       & 0.613 & 0.112 & 0.200 & 0.892 \\
        $PE_{rep}$(residential)  & 0.541 & 0.166 & 0.057 & 0.958 \\
        $PE_{dem}$(offline)      & 0.334 & 0.133 & 0.047 & 0.834 \\
        $PE_{dem}$(online)       & 0.309 & 0.107 & 0.069 & 0.778 \\
        $PE_{dem}$(residential)  & 0.349 & 0.168 & 0.021 & 0.925 \\
        \hline
    \end{tabular}
    }
    \label{tab:statistics_exposure}
\end{table}

% t-test tra dimensioni
\setlength{\tabcolsep}{20pt} 
\renewcommand{\arraystretch}{1} 
\begin{table}[h]
\resizebox{\textwidth}{!}{%
\begin{tabular}{@{}lll|rr|rr@{}}
\textbf{Exposure to} & \textbf{Dimension 1} & \textbf{Dimension 2}  & \textbf{T-statistic} & \textbf{Significance} & \textbf{KS statistic}  & \textbf{Significance} \\ \midrule
DEM       & Offline & Online                                   & 7.950 &  *** & 0.099 &  ***\\
DEM       & Offline & Residential                               & -3.968 &  *** & 0.072 &  ***\\ 
DEM       & Online  & Residential                                  &  -11.12 &  *** & 0.154 &  ***\\
\midrule
REP       & Offline & Online                                   & 1.931 & ns & 0.093 &  ***\\
REP       & Offline & Residential                              & 20.31 &  *** & 0.212 &  ***\\
REP       & Online  & Residential                                  & 20.22 &  *** & 0.230 &  ***\\ 
\bottomrule
\multicolumn{7}{@{}l@{}}{\footnotesize Note: $^{*}\, p<.05$; $^{**}\, p<.01$; $^{***}\, p<.001$}
\end{tabular}%
}
\caption{Comparison of the dimensions for all the counties in the contiguous United States using t-test for the statistical significance.}
\label{tab:t-test-dimensions}
\end{table}

% t-test tra dimension in metro areas
\setlength{\tabcolsep}{20pt} 
\renewcommand{\arraystretch}{1} 
\begin{table}[h]
\resizebox{\textwidth}{!}{%
\begin{tabular}{@{}lll|rr|rr@{}}
\textbf{Exposure to} & \textbf{Dimension 1} & \textbf{Dimension 2}  & \textbf{T-statistic} & \textbf{Significance} & \textbf{KS statistic}  & \textbf{Significance} \\ \midrule
DEM - Metro      & Offline & Online                            & 9.382 & *** & 0.171 & ***\\
DEM - Metro      & Offline & Residential                       & -0.991 &  ns & 0.067 & *\\ 
DEM - Metro      & Online  & Residential                           & -9.383 &  *** & 0.218 & ***\\
\midrule
REP - Metro      & Offline & Online                            & -2.205 & * & 0.122 & ***\\
REP - Metro      & Offline & Residential                       & 12.87 & *** & 0.234 & ***\\
REP - Metro      & Online  & Residential                           & 16.69 & *** & 0.324 & ***\\ 
\bottomrule
\multicolumn{7}{@{}l@{}}{\footnotesize Note: $^{*}\, p<.05$; $^{**}\, p<.01$; $^{***}\, p<.001$}
\end{tabular}%
}
\caption{Comparison of the dimensions in metropolitan areas using t-test for the statistical significance.}
\label{tab:t-test-dimensions-metro}
\end{table}

% t-test tra dimension in non-metro areas
\setlength{\tabcolsep}{20pt} 
\renewcommand{\arraystretch}{1} 
\begin{table}[htp]
\resizebox{\textwidth}{!}{%
\begin{tabular}{@{}lll|rr|rr@{}}
\textbf{Exposure to} & \textbf{Dimension 1} & \textbf{Dimension 2}  & \textbf{T-statistic} & \textbf{Significance} & \textbf{KS statistic}  & \textbf{Significance} \\ \midrule
DEM - Non-Metro      & Offline & Online                            & 3.438 & *** & 0.060 & **\\
DEM - Non-Metro      & Offline & Residential                       & -4.287 & *** & 0.844 & ***\\ 
DEM - Non-Metro      & Online  & Residential                           &  -7.351 & *** & 0.119 &\\
\midrule
REP - Non-Metro      & Offline & Online                            & 4.011 & *** & 0.102 & *** \\
REP - Non-Metro      & Offline & Residential                       & 16.69 & *** & 0.237 & *** \\
REP - Non-Metro      & Online  & Residential                           & 13.96 & *** & 0.186 & *** \\ 
\bottomrule
\multicolumn{7}{@{}l@{}}{\footnotesize Note: $^{*}\, p<.05$; $^{**}\, p<.01$; $^{***}\, p<.001$}
\end{tabular}%
}
\caption{Comparison of the dimensions in non-metropolitan areas using t-statistic for the statistical significance.}
\label{tab:t-test-dimensions-non-metro}
\end{table}

\setlength{\tabcolsep}{20pt} 
\renewcommand{\arraystretch}{1} 
\begin{table}[htp]
\resizebox{\textwidth}{!}{%
\begin{tabular}{@{}lll|rr|rr@{}}
\textbf{Exposure to} & \textbf{Dimension 1} & \textbf{Dimension 2}  & \textbf{T-statistic} & \textbf{Significance} & \textbf{KS statistic}  & \textbf{Significance} \\ \midrule
DEM      & Offline - Metro & Offline - Non-Metro                & 15.92 & *** & 0.279 & *** \\
DEM      & Online - Metro & Online - Non-Metro                  & 12.47 & *** & 0.256 & ***\\ 
DEM      & Residential - Metro & Residential - Non-Metro        & 10.20 & *** & 0.226 & ***\\
\midrule
REP      & Offline - Metro & Offline - Non-Metro                & -17.72 & *** & 0.294 & ***\\
REP      & Online - Metro & Online - Non-Metro                  & -16.08 & *** & 0.301 & ***\\ 
REP      & Residential - Metro & Residential - Non-Metro        & -14.06 & *** & 0.239 & ***\\
\bottomrule
\multicolumn{7}{@{}l@{}}{\footnotesize Note: $^{*}\, p<.05$; $^{**}\, p<.01$; $^{***}\, p<.001$}
\end{tabular}%
}
\caption{Comparison between exposure in metropolitan and non-metropolitan areas using Welch's t-test for the statistical significance.}
\label{tab:t-test-metro-non}
\end{table}

\clearpage
\section{Partisan segregation results}
As explained in the manuscript, we enhance model explainability related to the prediction of physical, online, and residential partisan segregation by computing the SHAP values. The absolute impact of the predictors on the final predictions are the following.

\setlength{\tabcolsep}{40pt} 
\renewcommand{\arraystretch}{1} 
\begin{table}[htp]
\resizebox{\textwidth}{!}{%
\begin{tabular}{@{}lrrr@{}}
\toprule
\textbf{Variable} & \textbf{Offline} & \textbf{Online}  & \textbf{Residential} \\ \midrule
\% graduated              &  0.095842  &   0.067271   &   0.067291     \\
\% urban population       &  0.052936  &   0.022199   &   0.080493     \\ 
\% unemployed             &  0.048427  &  0.034916    &   0.056921     \\ 
\% African Americans      &  0.037656  &   0.037174   &   0.074869     \\
\% Latinos/Hispanics      &  0.021695  &  0.016638    &   0.028644     \\ 
\bottomrule
\end{tabular}%
}
\caption{Absolute impacts of the predictors on the final prediction (ordered according to the offline prediction).}
\label{tab:shap}
\end{table}

\clearpage
\section{Regression results}

\subsection{Spatial autoregressive lag models}

We model the relationship between partisan exposure and voting patterns of all the counties of the contiguous United States with spatial autoregressive lag models \cite{ans88}. We compute spatial weights using k-nearest neighbour, with $k$ equal to $5$, $7$, and $10$. The variables are standardized. The results are the following. 

\setlength{\tabcolsep}{16pt} 
\begin{table}[ht]
  \centering
  \caption{Spatial autoregressive lag model with k=5 \\ Relationship between partisan exposure and voting patterns of US counties} 
  \label{}
  \resizebox{\textwidth}{!}{%
  \begin{tabular}{@{} p{3cm} c c c c c c @{}} 
  \toprule
  & \multicolumn{3}{c}{Share of Republican votes} & \multicolumn{3}{c}{Share of Democratic votes} \\ 
  \cmidrule(lr){2-4} \cmidrule(lr){5-7}
  & \multicolumn{1}{p{1.5cm}}{\centering Offline} & \multicolumn{1}{p{1.5cm}}{\centering Online} & \multicolumn{1}{p{1.5cm}}{\centering Residential} & \multicolumn{1}{p{1.5cm}}{\centering Offline} & \multicolumn{1}{p{1.5cm}}{\centering Online} & \multicolumn{1}{p{1.5cm}}{\centering Residential} \\  
  \midrule
   $\rho$ & $-0.204^{***}$ & $-0.030$ & $0.464^{***}$ & $-0.201^{***}$ & $-0.038^{*}$ & $0.562^{***}$ \\ 
    & $(0.007)$ &  $(0.016)$ &  $(0.013)$ & $(0.007)$ & $(0.016)$ & $(0.013)$ \\ [1ex]
   Intercept & $0.000$ & $0.000$ & $0.000$ & $0.000$ & $0.000$ & $0.000$ \\ 
    & $(0.003)$ &  $(0.007)$ &  $(0.008)$ & $(0.003)$ & $(0.007)$ & $(0.009)$ \\ [1ex]
   $PE_{rep}$ Offline & $1.121^{***}$ &  &  &  &  &  \\ 
    &  $(0.006)$ &  &  &  &  &  \\  [1ex]
   $PE_{rep}$ Online &  & $0.945^{***}$ &  &  &  &  \\ 
    &  &  $(0.014)$ &  &  &  &  \\ [1ex]
   $PE_{rep}$ Residential&  & & $0.598^{***}$ &  &  &  \\ 
    &  &  & $(0.012)$ &  &  &  \\ [1ex]
   $PE_{dem}$ Offline & &  &  & $1.118^{***}$ &  &  \\ 
    &   &  &  & $(0.006)$ &  &  \\  [1ex]
   $PE_{dem}$ Online &  &  &  &  & $0.956^{***}$ &  \\ 
    &  &   &  &  & $(0.013)$ &  \\ [1ex]
   $PE_{dem}$ Residential &  & &  &  &  & $0.500^{***}$ \\ 
    &  &  &  &  &  & $(0.013)$ \\ [1ex]
  \midrule 
     $R^2$ & $0.971$ & $0.854$ & $0.803$ & $0.967$ & $0.866$ & $0.754$ \\
     $Log-Likelihood$ & $1085.806$ & $-1414.123$ & $-1947.910$ & $891.834$ & $-1284.510$ & $-2333.192$ \\
     $AIC$ & $-2163.612$ &  $2836.247$ & $3903.820$ & $-1775.667$ & $2577.019$ & $4674.384$ \\
     N & $3098$ & $3098$ & $3098$ & $3098$ & $3098$ & $3098$ \\
  \bottomrule
  \multicolumn{7}{@{}l@{}}{\footnotesize Note: $^{*}\, p<.05$; $^{**}\, p<.01$; $^{***}\, p<.001$}
  \end{tabular}
  }
\end{table}

\begin{table}[ht]
  \centering
  \caption{Spatial autoregressive lag model with k=7 \\ Relationship between partisan exposure and voting patterns of US counties} 
  \label{}
  \resizebox{\textwidth}{!}{%
  \begin{tabular}{@{} p{3cm} c c c c c c @{}} 
  \toprule
  & \multicolumn{3}{c}{Share of Republican votes} & \multicolumn{3}{c}{Share of Democratic votes} \\ 
  \cmidrule(lr){2-4} \cmidrule(lr){5-7}
  & \multicolumn{1}{p{1.5cm}}{\centering Offline} & \multicolumn{1}{p{1.5cm}}{\centering Online} & \multicolumn{1}{p{1.5cm}}{\centering Residential} & \multicolumn{1}{p{1.5cm}}{\centering Offline} & \multicolumn{1}{p{1.5cm}}{\centering Online} & \multicolumn{1}{p{1.5cm}}{\centering Residential} \\  
  \midrule
   $\rho$ & $-0.216^{***}$ & $-0.051^{**}$ & $0.485^{***}$ & $-0.211^{***}$ & $-0.055^{***}$ & $0.588^{***}$ \\ 
    & $(0.007)$ &  $(0.017)$ &  $(0.013)$ & $(0.008)$ & $(0.016)$ & $(0.013)$ \\ [1ex]
   Intercept & $-0.001$ & $0.000$ & $0.002$ & $0.001$ & $0.000$ & $-0.002$ \\ 
    & $(0.003)$ &  $(0.007)$ &  $(0.008)$ & $(0.003)$ & $(0.007)$ & $(0.009)$ \\ [1ex]
   $PE_{rep}$ Offline & $1.125^{***}$ &  &  &  &  &  \\ 
    &  $(0.006)$ &  &  &  &  &  \\  [1ex]
   $PE_{rep}$ Online &  & $0.959^{***}$ &  &  &  &  \\ 
    &  &  $(0.014)$ &  &  &  &  \\ [1ex]
   $PE_{rep}$ Residential&  & & $0.595^{***}$ &  &  &  \\ 
    &  &  & $(0.012)$ &  &  &  \\ [1ex]
   $PE_{dem}$ Offline & &  &  & $1.120^{***}$ &  &  \\ 
    &   &  &  & $(0.006)$ &  &  \\  [1ex]
   $PE_{dem}$ Online &  &  &  &  & $0.967^{***}$ &  \\ 
    &  &   &  &  & $(0.013)$ &  \\ [1ex]
   $PE_{dem}$ Residential &  & &  &  &  & $0.496^{***}$ \\ 
    &  &  &  &  &  & $(0.013)$ \\ [1ex]
  \midrule 
     $R^2$ & $0.972$ & $0.854$ & $0.803$ & $0.968$ & $0.866$ & $0.754$ \\
     $Log-Likelihood$ & $1108.953$ & $-1410.881$ & $-1933.418$ & $904.876$ & $-1281.338$ & $-2312.850$ \\
     $AIC$ & $-2209.905$ &  $2829.762$ & $3874.836$ & $-1801.752$ & $2570.677$ & $4633.700$ \\
     N & $3098$ & $3098$ & $3098$ & $3098$ & $3098$ & $3098$ \\
  \bottomrule
  \multicolumn{7}{@{}l@{}}{\footnotesize Note: $^{*}\, p<.05$; $^{**}\, p<.01$; $^{***}\, p<.001$}
  \end{tabular}
  }
\end{table}

\begin{table}[ht]
  \centering
  \caption{Spatial autoregressive lag model with k=10 \\ Relationship between partisan exposure and voting patterns of US counties} 
  \label{}
  \resizebox{\textwidth}{!}{%
  \begin{tabular}{@{} p{3cm} c c c c c c @{}} 
  \toprule
  & \multicolumn{3}{c}{Share of Republican votes} & \multicolumn{3}{c}{Share of Democratic votes} \\ 
  \cmidrule(lr){2-4} \cmidrule(lr){5-7}
  & \multicolumn{1}{p{1.5cm}}{\centering Offline} & \multicolumn{1}{p{1.5cm}}{\centering Online} & \multicolumn{1}{p{1.5cm}}{\centering Residential} & \multicolumn{1}{p{1.5cm}}{\centering Offline} & \multicolumn{1}{p{1.5cm}}{\centering Online} & \multicolumn{1}{p{1.5cm}}{\centering Residential} \\  
  \midrule
   $\rho$ & $-0.230^{***}$ & $-0.090^{***}$ & $0.504^{***}$ & $-0.228^{***}$ & $-0.088^{*}$ & $0.613^{***}$ \\ 
    & $(0.007)$ &  $(0.017)$ &  $(0.013)$ & $(0.008)$ & $(0.017)$ & $(0.013)$ \\ [1ex]
   Intercept & $-0.001$ & $0.000$ & $0.003$ & $0.001$ & $0.000$ & $-0.003$ \\ 
    & $(0.003)$ &  $(0.007)$ &  $(0.008)$ & $(0.003)$ & $(0.007)$ & $(0.009)$ \\ [1ex]
   $PE_{rep}$ Offline & $1.129^{***}$ &  &  &  &  &  \\ 
    &  $(0.006)$ &  &  &  &  &  \\  [1ex]
   $PE_{rep}$ Online &  & $0.983^{***}$ &  &  &  &  \\ 
    &  &  $(0.014)$ &  &  &  &  \\ [1ex]
   $PE_{rep}$ Residential&  & & $0.597^{***}$ &  &  &  \\ 
    &  &  & $(0.011)$ &  &  &  \\ [1ex]
   $PE_{dem}$ Offline & &  &  & $1.126^{***}$ &  &  \\ 
    &   &  &  & $(0.006)$ &  &  \\  [1ex]
   $PE_{dem}$ Online &  &  &  &  & $0.988^{***}$ &  \\ 
    &  &   &  &  & $(0.013)$ &  \\ [1ex]
   $PE_{dem}$ Residential &  & &  &  &  & $0.498^{***}$ \\ 
    &  &  &  &  &  & $(0.013)$ \\ [1ex]
  \midrule 
     $R^2$ & $0.972$ & $0.855$ & $0.803$ & $0.968$ & $0.867$ & $0.752$ \\
     $Log-Likelihood$ & $1145.997$ & $-1401.403$ & $-1926.848$ & $945.742$ & $-1272.408$ & $-2307.772$ \\
     $AIC$ & $-2283.994$ &  $2810.805$ & $3861.696$ & $-1883.484$ & $2552.815$ & $4623.545$ \\
     N & $3098$ & $3098$ & $3098$ & $3098$ & $3098$ & $3098$ \\
  \bottomrule
  \multicolumn{7}{@{}l@{}}{\footnotesize Note: $^{*}\, p<.05$; $^{**}\, p<.01$; $^{***}\, p<.001$}
  \end{tabular}
  }
\end{table}

\clearpage
\subsection{Direct, indirect, and total effects in spatial lag models}
We compute the direct, indirect, and total effects of the independent variables for spatial autoregressive lag models with $k=7$. Table \ref{tab:effects} outlines the decomposition of these effects.

\setlength{\tabcolsep}{24pt}
\renewcommand{\arraystretch}{1} 
\begin{table}[h]
    \centering
    \caption{Direct, indirect, and total effects of the independent variables, with standard errors and statistical significance.}
    \resizebox{\textwidth}{!}{%
    \begin{tabular}{@{}lcrrr@{}}
        \hline
        Exposure & Vote & Direct & Indirect & Total \\
        \hline
        $PE_{rep}$(offline)      & Republican & 1.131*** (0.006) & -0.206*** (0.007) & 0.925*** (0.003) \\
        $PE_{rep}$(online)       & Republican & 0.959*** (0.014) & -0.047**  (0.015) & 0.912*** (0.008)\\
        $PE_{rep}$(residential)  & Republican & 0.619*** (0.011) & 0.536***  (0.022) & 1.155*** (0.021)\\
        $PE_{dem}$(offline)      & Democrat   & 1.126*** (0.006) & -0.201*** (0.007) & 0.925*** (0.003)\\
        $PE_{dem}$(online)       & Democrat   & 0.968*** (0.013) & -0.051*** (0.015) & 0.917*** (0.007)\\
        $PE_{dem}$(residential)  & Democrat   & 0.530*** (0.012) & 0.676*** (0.027) & 1.206*** (0.028) \\
        \hline
    \end{tabular}
    }
    \label{tab:effects}
\end{table}

\clearpage
\subsection{OLS models for metro and non-metro areas}
We analyse the relationship between partisan exposure and voting patterns for both metropolitan and non-metro areas employing Ordinary Least Squares (OLS) regressions. The variables are standardized. The complete results are the following.

\setlength{\tabcolsep}{16pt}
\begin{table}[ht]
  \centering
  \caption{OLS models for metropolitan areas \\ Relationship between partisan exposure and voting patterns} 
  \label{}
  \resizebox{\textwidth}{!}{%
  \begin{tabular}{@{} p{3cm} c c c c c c @{}} 
  \toprule
  & \multicolumn{3}{c}{Share of Republican votes} & \multicolumn{3}{c}{Share of Democratic votes} \\ 
  \cmidrule(lr){2-4} \cmidrule(lr){5-7}
  & \multicolumn{1}{p{1.5cm}}{\centering Offline} & \multicolumn{1}{p{1.5cm}}{\centering Online} & \multicolumn{1}{p{1.5cm}}{\centering Residential} & \multicolumn{1}{p{1.5cm}}{\centering Offline} & \multicolumn{1}{p{1.5cm}}{\centering Online} & \multicolumn{1}{p{1.5cm}}{\centering Residential} \\  
  \midrule
   Intercept & $0.000$ & $0.000$ & $0.000$ & $0.000$ & $0.000$ & $0.000$ \\ 
    & $(0.008)$ &  $(0.015)$ &  $(0.013)$ & $(0.009)$ & $(0.014)$ & $(0.016)$ \\ [1ex]
   $PE_{rep}$ Offline & $0.963^{***}$ &  &  &  &  &  \\ 
    &  $(0.008)$ &  &  &  &  &  \\  [1ex]
   $PE_{rep}$ Online &  & $0.853^{***}$ &  &  &  &  \\ 
    &  &  $(0.015)$ &  &  &  &  \\ [1ex]
   $PE_{rep}$ Residential&  & & $0.894^{***}$ &  &  &  \\ 
    &  &  & $(0.013)$ &  &  &  \\ [1ex]
   $PE_{dem}$ Offline & &  &  & $0.957^{***}$ &  &  \\ 
    &   &  &  & $(0.009)$ &  &  \\  [1ex]
   $PE_{dem}$ Online &  &  &  &  & $0.873^{***}$ &  \\ 
    &  &   &  &  & $(0.014)$ &  \\ [1ex]
   $PE_{dem}$ Residential &  & &  &  &  & $0.833^{***}$ \\ 
    &  &  &  &  &  & $(0.018)$ \\ [1ex]
  \midrule 
     $R^2$ & $0.927$ & $0.727$ & $0.800$ & $0.916$ & $0.763$ & $0.693$ \\
     N & $1156$ & $1156$ & $1156$ & $1156$ & $1156$ & $1156$ \\
  \bottomrule
  \multicolumn{7}{@{}l@{}}{\footnotesize Note: $^{*}\, p<.05$; $^{**}\, p<.01$; $^{***}\, p<.001$}
  \end{tabular}
  }
\end{table}

\begin{table}[ht]
  \centering
  \caption{OLS models for non-metro areas \\ Relationship between partisan exposure and voting patterns} 
  \label{}
  \resizebox{\textwidth}{!}{%
  \begin{tabular}{@{} p{3cm} c c c c c c @{}} 
  \toprule
  & \multicolumn{3}{c}{Share of Republican votes} & \multicolumn{3}{c}{Share of Democratic votes} \\ 
  \cmidrule(lr){2-4} \cmidrule(lr){5-7}
  & \multicolumn{1}{p{1.5cm}}{\centering Offline} & \multicolumn{1}{p{1.5cm}}{\centering Online} & \multicolumn{1}{p{1.5cm}}{\centering Residential} & \multicolumn{1}{p{1.5cm}}{\centering Offline} & \multicolumn{1}{p{1.5cm}}{\centering Online} & \multicolumn{1}{p{1.5cm}}{\centering Residential} \\  
  \midrule
   Intercept & $0.000$ & $0.000$ & $0.000$ & $0.000$ & $0.000$ & $0.000$ \\ 
    & $(0.003)$ &  $(0.006)$ &  $(0.014)$ & $(0.003)$ & $(0.005)$ & $(0.016)$ \\ [1ex]
   $PE_{rep}$ Offline & $0.991^{***}$ &  &  &  &  &  \\ 
    &  $(0.003)$ &  &  &  &  &  \\  [1ex]
   $PE_{rep}$ Online &  & $0.969^{***}$ &  &  &  &  \\ 
    &  &  $(0.006)$ &  &  &  &  \\ [1ex]
   $PE_{rep}$ Residential&  & & $0.790^{***}$ &  &  &  \\ 
    &  &  & $(0.014)$ &  &  &  \\ [1ex]
   $PE_{dem}$ Offline & &  &  & $0.993^{***}$ &  &  \\ 
    &   &  &  & $(0.003)$ &  &  \\  [1ex]
   $PE_{dem}$ Online &  &  &  &  & $0.973^{***}$ &  \\ 
    &  &   &  &  & $(0.005)$ &  \\ [1ex]
   $PE_{dem}$ Residential &  & &  &  &  & $0.705^{***}$ \\ 
    &  &  &  &  &  & $(0.016)$ \\ [1ex]
  \midrule 
     $R^2$ & $0.983$ & $0.939$ & $0.623$ & $0.985$ & $0.947$ & $0.497$ \\
     N & $1942$ & $1942$ & $1942$ & $1942$ & $1942$ & $1942$ \\
  \bottomrule
  \multicolumn{7}{@{}l@{}}{\footnotesize Note: $^{*}\, p<.05$; $^{**}\, p<.01$; $^{***}\, p<.001$}
  \end{tabular}
  }
\end{table}

\clearpage
\subsection{Marginal effects of spatial and OLS models}
\begin{figure}[htp]
    \centering
    \includegraphics[width=1\linewidth]{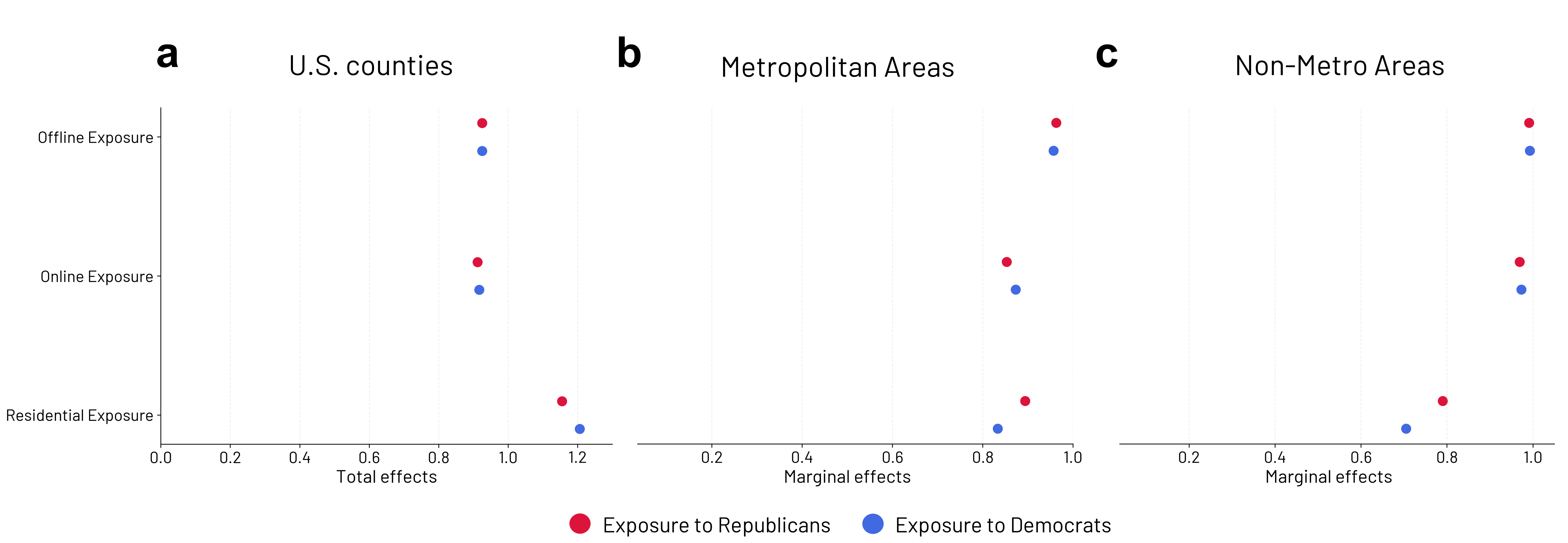}
    \caption{\textbf{Total and marginal effects of spatial and OLS models.} }
    \label{SI-fig-effects}
\end{figure}

Fig. \ref{SI-fig-effects} shows the total effects of the spatial autoregressive lag models, using spatial weights with $k=7$, and the marginal effects of the OLS regressions for metropolitan and non-metropolitan areas. All variables have been standardized, so the coefficients can be interpreted as the change in the dependent variable associated with a one-standard-deviation increase in the independent variable.

\clearpage
\section{Dominance analysis results}
We enhance the robustness of the results by employing dominance analysis. The results are the following.

\setlength{\tabcolsep}{15pt}
\renewcommand{\arraystretch}{1}
\begin{table}[htp]
\resizebox{\textwidth}{!}{%
\begin{tabular}{@{}lllllr@{}}
\toprule
\textbf{Variable} & \makecell{\textbf{Interactional} \\ \textbf{Dominance}} & \makecell{\textbf{Individual} \\ \textbf{Dominance}}  & \makecell{\textbf{Average} \\ \textbf{Partial} \\ \textbf{Dominance}} & \makecell{\textbf{Total} \\ \textbf{Dominance}} & \makecell{\textbf{Percentage} \\ \textbf{Relative} \\ \textbf{Importance}} \\ \midrule

Physical exposure          & 0.052839 & 0.962836 & 0.287563 & 0.342631 & 35.287168 \\
Online exposure            & 0.00135  & 0.853889 & 0.212643 & 0.266387 & 27.434839 \\
Residential exposure       & 0.002551 & 0.697881 & 0.143587 & 0.195244 & 20.107934 \\
\midrule
\% graduated               & 0.00048  & 0.209583 & 0.040195 & 0.056404 & 5.808968  \\
\% urban population        & 0.000277 & 0.213559 & 0.026400 & 0.04653  & 4.792054  \\
\% African Americans       & 0.000844 & 0.176048 & 0.027653 & 0.042851 & 4.413207  \\
\% unemployed              & 0.000015 & 0.063053 & 0.010888 & 0.016049 & 1.652886  \\
\% Latinos/Hispanics       & 0.000033 & 0.017231 & 0.003634 & 0.004883 & 0.502944  \\
\bottomrule
\end{tabular}%
}
\caption{Dominance Analysis of the relationship between partisan exposure to Republicans (and demographic and socioeconomic characteristics) and votes for the Republican party.}
\label{tab:dominance-rep}
\end{table}

\setlength{\tabcolsep}{15pt} 
\renewcommand{\arraystretch}{1} 
\begin{table}[htp]
\resizebox{\textwidth}{!}{%
\begin{tabular}{@{}lllllr@{}}
\toprule
\textbf{Variable} & \makecell{\textbf{Interactional} \\ \textbf{Dominance}} & \makecell{\textbf{Individual} \\ \textbf{Dominance}}  & \makecell{\textbf{Average} \\ \textbf{Partial} \\ \textbf{Dominance}} & \makecell{\textbf{Total} \\ \textbf{Dominance}} & \makecell{\textbf{Percentage} \\ \textbf{Relative} \\ \textbf{Importance}} \\ \midrule

Physical exposure          & 0.044584 & 0.959156 & 0.299083 & 0.34978  & 36.120709 \\
Online exposure            & 0.001604 & 0.865488 & 0.236315 & 0.285623 & 29.495423 \\
Residential exposure       & 0.001567 & 0.570881 & 0.110852 & 0.154695 & 15.974911 \\
\midrule
\% graduated               & 0.000479 & 0.182337 & 0.040945 & 0.05356  & 5.531018  \\
\% African Americans       & 0.000795 & 0.217876 & 0.033931 & 0.052782 & 5.450677  \\
\% urban population        & 0.000917 & 0.200255 & 0.031387 & 0.048687 & 5.027732  \\
\% unemployed              & 0.000012 & 0.074862 & 0.012022 & 0.018375 & 1.897572  \\
\% Latinos/Hispanics       & 0.000011 & 0.017489 & 0.003564 & 0.004861 & 0.501957  \\
\bottomrule
\end{tabular}%
}
\caption{Dominance Analysis of the relationship between partisan exposure to Democrats (and demographic and socioeconomic characteristics) and votes for the Democratic party.}
\label{tab:dominance-dem}
\end{table}

\clearpage
\subsection{Dominance analysis results for metro and non-metro areas}

\setlength{\tabcolsep}{15pt} 
\renewcommand{\arraystretch}{1} 
\begin{table}[htp]
\resizebox{\textwidth}{!}{%
\begin{tabular}{@{}lllllr@{}}
\toprule
\textbf{Variable} & \makecell{\textbf{Interactional} \\ \textbf{Dominance}} & \makecell{\textbf{Individual} \\ \textbf{Dominance}}  & \makecell{\textbf{Average} \\ \textbf{Partial} \\ \textbf{Dominance}} & \makecell{\textbf{Total} \\ \textbf{Dominance}} & \makecell{\textbf{Percentage} \\ \textbf{Relative} \\ \textbf{Importance}} \\ \midrule

Physical exposure          & 0.052123 & 0.927305 & 0.251718 & 0.311217 & 32.578675 \\
Residential exposure       & 0.011786 & 0.799454 & 0.174116 & 0.231992 & 24.285288 \\
Online exposure            & 0.001519 & 0.726874 & 0.146281 & 0.20076  & 21.015822 \\
\midrule
\% graduated               & 0.000827 & 0.297376 & 0.049967 & 0.074751 & 7.82503  \\
\% urban population        & 0.001369 & 0.326641 & 0.039834 & 0.070877 & 7.419502  \\
\% African Americans       & 0.00166  & 0.145012 & 0.022990 & 0.035577 & 3.72424  \\
\% unemployed              & 0.000039 & 0.060931 & 0.010337 & 0.015374 & 1.609367  \\
\% Latinos/Hispanics       & 0.0      & 0.060495 & 0.009559 & 0.014731 & 1.542076  \\
\bottomrule
\end{tabular}%
}
\caption{Dominance Analysis of the relationship between partisan exposure to Republicans (and demographic and socioeconomic characteristics) and votes for the Republican party in metropolitan areas.}
\label{tab:dominance-metro-rep}
\end{table}

\setlength{\tabcolsep}{15pt} 
\renewcommand{\arraystretch}{1} 
\begin{table}[htp]
\resizebox{\textwidth}{!}{%
\begin{tabular}{@{}lllllr@{}}
\toprule
\textbf{Variable} & \makecell{\textbf{Interactional} \\ \textbf{Dominance}} & \makecell{\textbf{Individual} \\ \textbf{Dominance}}  & \makecell{\textbf{Average} \\ \textbf{Partial} \\ \textbf{Dominance}} & \makecell{\textbf{Total} \\ \textbf{Dominance}} & \makecell{\textbf{Percentage} \\ \textbf{Relative} \\ \textbf{Importance}} \\ \midrule

Physical exposure          & 0.051659 & 0.916374 & 0.254189 & 0.311646 & 32.914846 \\
Online exposure            & 0.002985 & 0.762566 & 0.165927 & 0.220139 & 23.250229 \\
Residential exposure       & 0.007668 & 0.693444 & 0.141521 & 0.19378  & 20.466278 \\
\midrule
\% graduated               & 0.00087  & 0.270723 & 0.051282 & 0.072411 & 7.647734  \\
\% urban population        & 0.003701 & 0.31066  & 0.043836 & 0.072172 & 7.622512  \\
\% African Americans       & 0.001767 & 0.180305 & 0.028092 & 0.043828 & 4.628951  \\
\% unemployed              & 0.000067 & 0.073395 & 0.011662 & 0.01793  & 1.893645  \\
\% Latinos/Hispanics       & 0.000032 & 0.060556 & 0.009795 & 0.01492  & 1.575805  \\
\bottomrule
\end{tabular}%
}
\caption{Dominance Analysis of the relationship between partisan exposure to Democrats (and demographic and socioeconomic characteristics) and votes for the Democratic party in metropolitan areas.}
\label{tab:dominance-metro-dem}
\end{table}

\clearpage

\setlength{\tabcolsep}{15pt} 
\renewcommand{\arraystretch}{1} 
\begin{table}[htp]
\resizebox{\textwidth}{!}{%
\begin{tabular}{@{}lllllr@{}}
\toprule
\textbf{Variable} & \makecell{\textbf{Interactional} \\ \textbf{Dominance}} & \makecell{\textbf{Individual} \\ \textbf{Dominance}}  & \makecell{\textbf{Average} \\ \textbf{Partial} \\ \textbf{Dominance}} & \makecell{\textbf{Total} \\ \textbf{Dominance}} & \makecell{\textbf{Percentage} \\ \textbf{Relative} \\ \textbf{Importance}} \\ \midrule

Physical exposure          & 0.033474 & 0.982865 & 0.322241 & 0.368723 & 37.394926 \\
Online exposure            & 0.001226 & 0.938518 & 0.284023 & 0.330485 & 33.516907 \\
Residential exposure       & 0.000508 & 0.623425 & 0.140635 & 0.183468 & 18.606826 \\
\midrule
\% African Americans       & 0.000238 & 0.181181 & 0.031178 & 0.046061 & 4.671355  \\
\% graduated               & 0.000274 & 0.060213 & 0.021464 & 0.023659 & 2.399414  \\
\% unemployed              & 0.000004 & 0.084427 & 0.012937 & 0.020257 & 2.054377  \\
\% urban population        & 0.000158 & 0.049613 & 0.007214 & 0.011632 & 1.179686  \\
\% Latinos/Hispanics       & 0.00009  & 0.003426 & 0.001735 & 0.00174  & 0.17651  \\
\bottomrule
\end{tabular}%
}
\caption{Dominance Analysis of the relationship between partisan exposure to Republicans (and demographic and socioeconomic characteristics) and votes for the Republican party in non-metro areas.}
\label{tab:dominance-nonmetro-rep}
\end{table}

\setlength{\tabcolsep}{15pt} 
\renewcommand{\arraystretch}{1} 
\begin{table}[htp]
\resizebox{\textwidth}{!}{%
\begin{tabular}{@{}lllllr@{}}
\toprule
\textbf{Variable} & \makecell{\textbf{Interactional} \\ \textbf{Dominance}} & \makecell{\textbf{Individual} \\ \textbf{Dominance}}  & \makecell{\textbf{Average} \\ \textbf{Partial} \\ \textbf{Dominance}} & \makecell{\textbf{Total} \\ \textbf{Dominance}} & \makecell{\textbf{Percentage} \\ \textbf{Relative} \\ \textbf{Importance}} \\ \midrule

Physical exposure          & 0.027059 & 0.985177 & 0.340459 & 0.381874 & 38.695702 \\
Online exposure            & 0.000725 & 0.9471   & 0.308820 & 0.350093 & 35.475296 \\
Residential exposure       & 0.000244 & 0.497242 & 0.104400 & 0.140486 & 14.235613 \\
\midrule
\% African Americans       & 0.00012  & 0.229318 & 0.039203 & 0.058082 & 5.885506  \\
\% unemployed              & 0.000009 & 0.09863  & 0.014312 & 0.023064 & 2.337096  \\
\% graduated               & 0.000156 & 0.040722 & 0.019669 & 0.019862 & 2.012615  \\
\% urban population        & 0.000066 & 0.042312 & 0.008817 & 0.01191  & 1.206865  \\
\% Latinos/Hispanics       & 0.000002 & 0.003612 & 0.001389 & 0.001493 & 0.151307  \\
\bottomrule
\end{tabular}%
}
\caption{Dominance Analysis of the relationship between partisan exposure to Democrats (and demographic and socioeconomic characteristics) and votes for the Democratic party in non-metro areas.}
\label{tab:dominance-nonmetro-dem}
\end{table}

\clearpage
\section{Random Forest and Elastic Net models}
We enhance the robustness and generalizability of the results by training Random Forest and Elastic Net \cite{zou2005regularization} models with k-fold cross-validation ($k = 5$), splitting the dataset into train and test sets with a 70/30 ratio. This analysis is performed considering all the counties of the contiguous United States ($N=3098$), for both Democrats and Republicans.

Physical partisan exposure is the most important predictor in the Random Forest models, as shown in Table \ref{tab:randomforest}.
The Elastic Net models achieve the best performances ($R^2=0.97$) with $\alpha = 1e-05$ and a ratio between L1 and L2 penalties of $0.1$, indicating that it is closer to an L2 (Ridge) penalty in both models. Table \ref{tab:elasticnet} shows the results of the beta coefficients.

\setlength{\tabcolsep}{50pt} 
\renewcommand{\arraystretch}{1}
\begin{table}[htp]
\resizebox{\textwidth}{!}{%
\begin{tabular}{@{}lrr@{}}
\textbf{Variable} & \textbf{Model 1 - Republicans} & \textbf{Model 2 - Democrats}  \\ \midrule
Physical exposure         &  0.9635  &    0.4731                  \\
Online exposure           &  0.0033  &    0.2805             \\ 
Residential exposure      &  0.0176  &    0.1516              \\
\% African Americans      &  0.0035  &    0.0208                   \\
\% graduated              &  0.0032  &    0.0259         \\
\% urban population       &  0.0058  &    0.0379               \\ 
\% unemployed             &  0.0014  &    0.0056              \\ 
\% Latinos/Hispanics      &  0.0017  &    0.0049              \\ 
\bottomrule
\end{tabular}%
}
\caption{Variable importance of the Random Forest models.}
\label{tab:randomforest}
\end{table}

\setlength{\tabcolsep}{50pt} 
\renewcommand{\arraystretch}{1} 
\begin{table}[htp]
\resizebox{\textwidth}{!}{%
\begin{tabular}{@{}lrr@{}}
\textbf{Variable} & \textbf{Model 1 - Republicans} & \textbf{Model 2 - Democrats}  \\ \midrule
Physical exposure         &  1.0819   &   1.1578                   \\
Online exposure           &  -0.1802  &   -0.2185               \\ 
Residential exposure      &  0.0906  &   0.0625                \\
\% African Americans      &  -0.0310  &   0.0322                    \\
\% graduated              &  -0.0539  &   0.0536               \\
\% urban population       &  -0.0100 &   0.0189                 \\ 
\% unemployed             &  0.0117  &  -0.0080                 \\ 
\% Latinos/Hispanics      &  -0.0058   &  -0.0027                 \\ 
\bottomrule
\end{tabular}
}
\caption{Beta coefficients of the Elastic Net models.}
\label{tab:elasticnet}
\end{table}

\clearpage
\section{Comparison between Colocation Maps and ACS Commuting Flows}
We compare the dataset used as a proxy for offline connectedness between US counties (Colocation Maps) with official statistics from the 2016-2020 5-Year ACS Commuting Flows dataset provided by the US Census Bureau. To this aim, we perform the same analyses employed to compare offline, online, and residential exposures.

Similar to the computation of co-location and friendship probabilities, we compute the commuting probability between two counties as the total flows between them divided by their population sizes. We then compute the partisan exposure as captured by the commuting flows and employ both spatial lag model and dominance analysis. We also perform t-tests to compare the difference between offline and commuting partisan exposures.

While the county-level distributions of partisan exposure to Democrats do not show significant differences (Table \ref{tab:t-test-commuting}), t-test and Kolmogorov-Smirnov tests reveal significant differences between offline and commuting exposure to Republicans ($P < .001$). Regarding the relationship between partisan exposure and voting patterns in US counties, partisan exposure as captured by commuting flows has slightly lower performance than offline exposure in explaining the variance of voting patterns, both in spatial models and dominance analysis.

\vspace{1cm}

\setlength{\tabcolsep}{15pt} 
\renewcommand{\arraystretch}{1} 
\begin{table}[h]
\resizebox{\textwidth}{!}{%
\begin{tabular}{@{}lll|rr|rr@{}}
\toprule
\textbf{Exposure to} & \textbf{Dimension 1} & \textbf{Dimension 2}  & \textbf{T-statistic} & \textbf{Significance} & \textbf{KS statistic}  & \textbf{Significance} \\ \midrule
DEM       & Offline & Commuting & -0.699 & ns  & 0.214 & ns\\
REP       & Offline & Commuting & -3.387 & *** & 0.054 & ***\\ 
\bottomrule
\multicolumn{7}{@{}l@{}}{\footnotesize Note: $^{*}\, p<.05$; $^{**}\, p<.01$; $^{***}\, p<.001$}
\end{tabular}%
}
\caption{Comparison between partisan exposure in offline and commuting networks using t-test for the statistical significance.}
\label{tab:t-test-commuting}
\end{table}

\begin{table}[ht]
  \centering
  \caption{Spatial autoregressive lag model with k=7 \\ Relationship between partisan exposure and voting patterns of US counties} 
  \label{tab:spatial_model}
  \scriptsize 
  \resizebox{15cm}{!}{%
  \begin{tabular}{@{} l c c c c c @{}} 
  \toprule
  & \multicolumn{2}{c}{Share of Republican votes} & \multicolumn{2}{c}{Share of Democratic votes} \\ 
  \cmidrule(lr){2-3} \cmidrule(lr){4-5}
  & \multicolumn{1}{c}{Offline} & \multicolumn{1}{c}{Commuting} & \multicolumn{1}{c}{Offline} & \multicolumn{1}{c}{Commuting} \\  
  \midrule
   $\rho$ & $-0.216^{***}$ & $-0.153^{***}$ & $-0.211^{***}$ & $-0.151^{***}$ \\ 
    & $(0.007)$ & $(0.011)$ & $(0.008)$ & $(0.011)$ \\ [0.5ex]
   Intercept & $-0.001$ & $-1.663^{***}$ & $0.001$ & $-0.03$ \\ 
    & $(0.003)$ & $(0.363)$ & $(0.003)$ & $(0.005)$ \\ [0.5ex]
   $PE_{rep}$ Offline & $1.125^{***}$ &  &  &  \\ 
    & $(0.006)$ &  &  &  \\  [0.5ex]
   $PE_{rep}$ Commuting &  & $1.066^{***}$ &  &  \\ 
    &  & $(0.008)$ &  &  \\ [0.5ex]
   $PE_{dem}$ Offline &  &  & $1.254^{***}$ &  \\ 
    &  &  & $(0.007)$ &  \\  [0.5ex]
   $PE_{dem}$ Commuting &  &  &  & $1.064^{***}$ \\ 
    &  &  &  & $(0.009)$ \\ [0.5ex]
  \midrule 
     $R^2$ & $0.972$ & $0.937$ & $0.968$ & $0.935$ \\
     $Log-Likelihood$ & $1108.953$ & $-125.143$ & $904.876$ & $-164.699$ \\
     $AIC$ & $-2209.905$ & $258.286$ & $-1801.752$ & $337.398$ \\
     N & $3098$ & $3090$ & $3098$ & $3090$ \\
  \bottomrule
  \multicolumn{5}{@{}l@{}}{\footnotesize Note: $^{*}\, p<.05$; $^{**}\, p<.01$; $^{***}\, p<.001$}
  \end{tabular}
  }
\end{table}

\setlength{\tabcolsep}{15pt} 
\renewcommand{\arraystretch}{1} 
\begin{table}[htp]
\resizebox{\textwidth}{!}{%
\begin{tabular}{@{}lllllr@{}}
\toprule
\textbf{Variable} & \makecell{\textbf{Interactional} \\ \textbf{Dominance}} & \makecell{\textbf{Individual} \\ \textbf{Dominance}}  & \makecell{\textbf{Average} \\ \textbf{Partial} \\ \textbf{Dominance}} & \makecell{\textbf{Total} \\ \textbf{Dominance}} & \makecell{\textbf{Percentage} \\ \textbf{Relative} \\ \textbf{Importance}} \\ \midrule

Physical exposure          & 0.020545 & 0.962702 & 0.186332 & 0.254175 & 26.17196 \\
Commuting exposure         & 0.0003   & 0.931946 & 0.161400 & 0.229117 & 23.591774 \\
Online exposure            & 0.001613 & 0.853406 & 0.135055 & 0.200045 & 20.598325 \\
Residential exposure       & 0.002468 & 0.697649 & 0.097064 & 0.153285 & 15.783504 \\
\midrule
\% graduated               & 0.000462 & 0.20697  & 0.027305 & 0.044285 & 4.559962  \\
\% urban population        & 0.000334 & 0.211635 & 0.019205 & 0.03849  & 3.963212  \\
\% African Americans       & 0.000862 & 0.177088 & 0.019570 & 0.034993 & 3.603222  \\
\% unemployed              & 0.000014 & 0.063017 & 0.007591 & 0.012907 & 1.329055  \\
\% Latinos/Hispanics       & 0.000029 & 0.01708  & 0.002538 & 0.003875 & 0.398987  \\
\bottomrule
\end{tabular}%
}
\caption{Dominance Analysis of the relationship between partisan exposure to Republicans (and demographic and socioeconomic characteristics) and votes for the Republican party by including commuting exposure.}
\label{tab:dominance-commuting-rep}
\end{table}

\setlength{\tabcolsep}{15pt} 
\renewcommand{\arraystretch}{1} 
\begin{table}[htp]
\resizebox{\textwidth}{!}{%
\begin{tabular}{@{}lllllr@{}}
\toprule
\textbf{Variable} & \makecell{\textbf{Interactional} \\ \textbf{Dominance}} & \makecell{\textbf{Individual} \\ \textbf{Dominance}}  & \makecell{\textbf{Average} \\ \textbf{Partial} \\ \textbf{Dominance}} & \makecell{\textbf{Total} \\ \textbf{Dominance}} & \makecell{\textbf{Percentage} \\ \textbf{Relative} \\ \textbf{Importance}} \\ \midrule

Physical exposure          & 0.020344 & 0.95901  & 0.192265 & 0.258356 & 26.67709 \\
Commuting exposure         & 0.000183 & 0.930383 & 0.167888 & 0.233976 & 24.159666 \\
Online exposure            & 0.001774 & 0.864938 & 0.147705 & 0.211183 & 21.806092 \\
Residential exposure       & 0.001593 & 0.570121 & 0.075455 & 0.122211 & 12.61917 \\
\midrule
\% African Americans       & 0.000834 & 0.219189 & 0.023897 & 0.043034 & 4.443513  \\
\% graduated               & 0.000482 & 0.179621 & 0.027598 & 0.041477 & 4.282771  \\
\% urban population        & 0.000856 & 0.198261 & 0.022310 & 0.039476 & 4.076222  \\
\% unemployed              & 0.000012 & 0.074858 & 0.008384 & 0.01484  & 1.532337  \\
\% Latinos/Hispanics       & 0.000005 & 0.017331 & 0.002543 & 0.003904 & 0.40314  \\
\bottomrule
\end{tabular}%
}
\caption{Dominance Analysis of the relationship between partisan exposure to Democrats (and demographic and socioeconomic characteristics) and votes for the Democratic party by including commuting exposure.}
\label{tab:dominance-commuting-dem}
\end{table}

\begin{figure}[htp]
    \centering
    \includegraphics[width=1\linewidth]{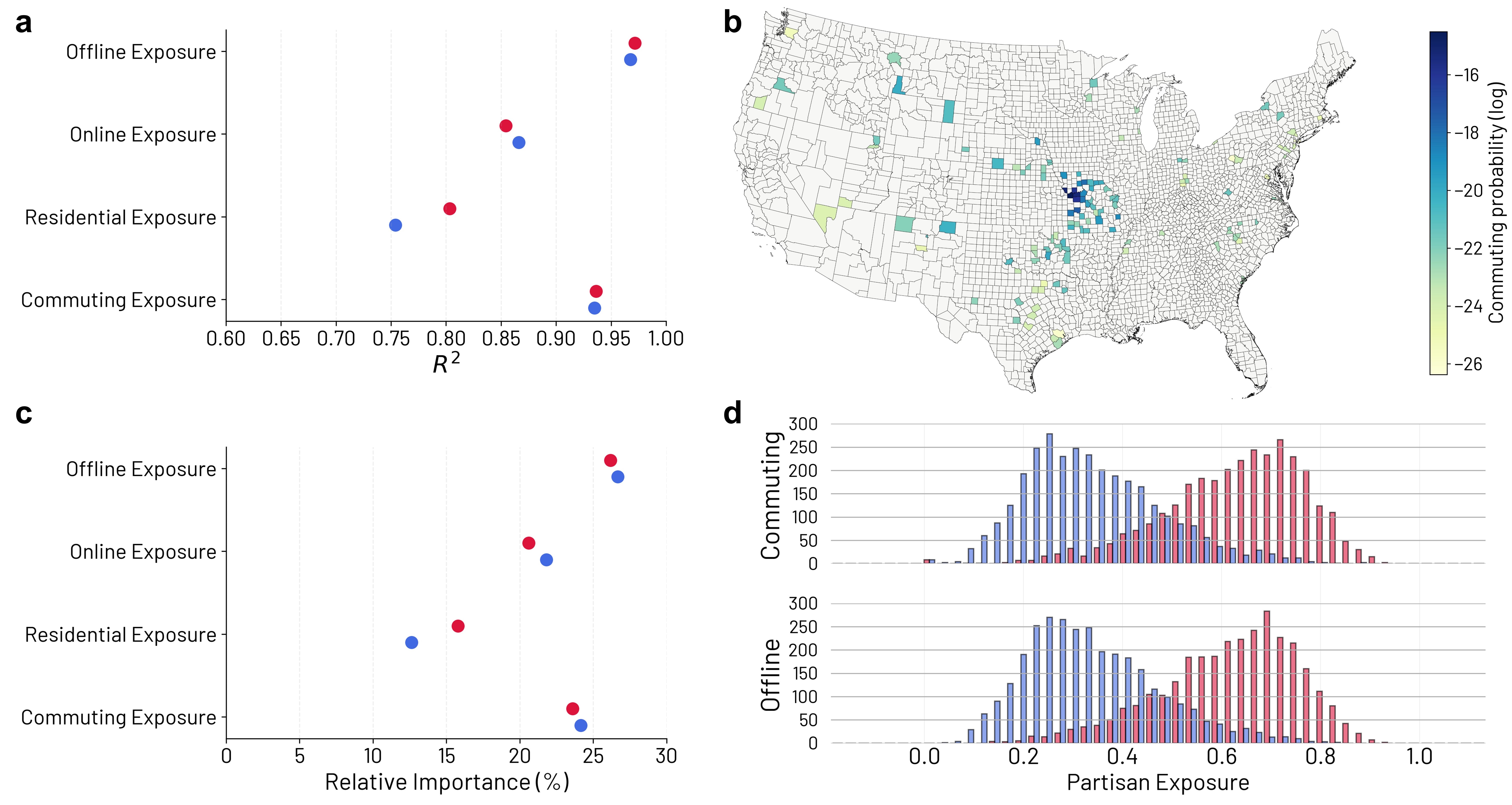}
    \caption{\textbf{Comparison between Colocation Maps and ACS Commuting Flows} 
    (\textbf{a}) $R^2$ of the spatial lag models. Commuting partisan exposure has a slightly lower $R^2$ than the offline one. (\textbf{b}) Map of the commuting probabilities (log) for the Jackson county. (\textbf{c}) Dominance analysis consider all the four dimensions. Commuting partisan exposure shows the lowest relative importance. (\textbf{d}) Distributions of offline and commuting partisan exposures to both Democrats and Republicans.}
    \label{SI-fig-commuting}
\end{figure}

\clearpage
\section{Exclude local exposure (self-loops) in the networks}
To achieve a comprehensive understanding of the relationship between partisan exposure and voting patterns, we perform all the analyses outlined in the paper excluding the self-loops from the co-location and online friendship networks. A self-loop refers to either the co-location or friendship probability between a county $i$ and itself. In the physical network, local exposure represents a substantial and fundamental aspect of an individual's exposure. Conversely, in the online network, while local exposure still has a huge impact, online exposure tends to be more heterogeneous and characterized by greater external exposure. 

\begin{figure}[htp]
    \centering
    \includegraphics[width=1\linewidth]{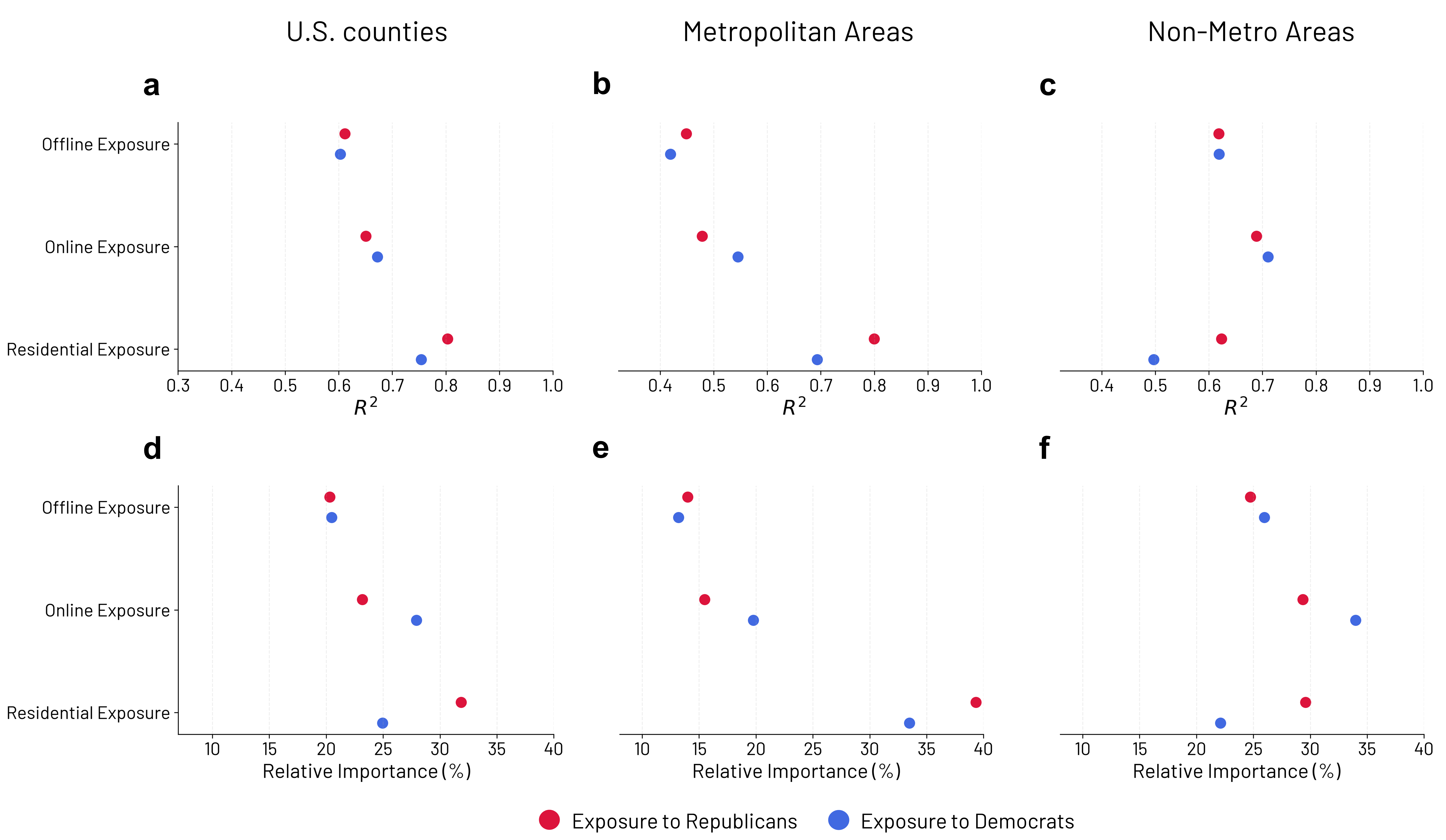}
    \caption{\textbf{Relative contribution of the three dimensions of partisan exposure on voting patterns, excluding local exposure in the networks.} 
    (\textbf{a}) $R^2$ of the spatial models with $k=7$ for all the counties in the contiguous United States. (\textbf{b}) $R^2$ of the OLS models for the metropolitan areas (RUCC 1--3). (\textbf{c}) $R^2$ of the OLS models for the non-metropolitan areas (RUCC 4--9). (\textbf{d}) Dominance analysis for all the US counties, considering the three dimensions, along with demographic and socioeconomic controls. (\textbf{e}) Dominance analysis for the metropolitan areas (RUCC 1--3). (\textbf{f}) Dominance analysis for the non-metropolitan areas (RUCC 4--9).}
    \label{SI-fig-noself}
\end{figure}

Excluding local exposure in the networks, physical proximity shows a significant loss of predictive and explanation power in determining US political outcomes at the county level (Fig. \ref{SI-fig-noself}). This is confirmed by both the spatial model (Fig. \ref{SI-fig-noself}a) and dominance analysis (Fig. \ref{SI-fig-noself}d) and both metropolitan (Fig. \ref{SI-fig-noself}b and \ref{SI-fig-noself}e) and non-metropolitan areas (Fig. \ref{SI-fig-noself}c and \ref{SI-fig-noself}f). Thus, online partisan exposure outweighs offline proximity in predicting US counties voting patterns, with greater differences between online and offline partisan exposure to Democrats.

\clearpage
\section{Descriptive and regression analyses of survey data}
We leverage the 2020-2022 Social Media Study provided by the American National Election Studies (ANES) to investigate the relationship between partisan exposure and voting behavior at the individual level. As explained in the Materials and Methods Section, we consider only those respondents who declared their vote preference in the second wave, have valid voting records, and responded to all the questions of interest. The final sample consists of 2,420 respondents, characterized as follows. 

\begin{table}[h]
    \centering
    \caption{Descriptive statistics of the analytical sample ($N=2420$)}
    \begin{tabular}{lcccccc}
        \hline
        Variable & Mean & SD & Min & Max\\
        \hline
        Gender & 1.50 & 0.50 & 1 & 2  \\
        Age & 50.59 & 15.88 & 18 & 80  \\
        Education - Graduation (dummy) & 0.45 & 0.50 & 0 & 1  \\
        Ethnicity: White  &   0.71 &  0.45 &  0  & 1  \\
        Ethnicity: Black   &  0.10  &  0.31 &  0 &  1  \\
        Ethnicity: Hispanic &  0.11  &  0.32 &  0  & 1   \\
        Employment & 2.67 & 2.14 & 1 & 7  \\
        Income & 10.60 & 4.02 & 1 & 18  \\
        W1 - Offline Exposure to Democrats & 3.08 & 1.08 & 1 & 5  \\
        W1 - Online Exposure to Democrats & 3.11 & 0.99 & 1 & 5\\
        W1 - Offline Exposure to Republicans & 2.83 & 1.04 & 1 & 5  \\
        W1 - Online Exposure to Republicans & 2.76 & 0.94 & 1 & 5  \\
        W2 - Offline Exposure to Democrats & 3.08 & 1.07 & 1 & 5  \\
        W1 - Online Exposure to Democrats & 3.05 & 1.06 & 1 & 5  \\
        W2 - Offline Exposure to Republicans & 2.85 & 1.04 & 1 & 5 \\
        W2 - Online Exposure to Republicans & 2.74 & 1.01 & 1 & 5  \\
        Political vote (0 Democrat - 1 Republican) & 0.42 & 0.49 & 0 & 1  \\
        \hline
    \end{tabular}
    \label{tab:descriptive_statistics}
\end{table}

\begin{table}[ht]
  \centering
  \caption{Logit models -- \\ Relationship between partisan exposure and political vote (0 Democrat - 1 Republican)} 
  \label{tab:spatial_model}
  \scriptsize % Reduce the font size to fit the table
  \resizebox{15cm}{!}{%
  \begin{tabular}{@{} l c c c c c @{}} 
  \toprule
  & \multicolumn{2}{c}{Exposure to Democrats} & \multicolumn{2}{c}{Exposure to Republicans} \\ 
  \cmidrule(lr){2-3} \cmidrule(lr){4-5}
  & \multicolumn{1}{c}{Wave 1} & \multicolumn{1}{c}{Wave 2} & \multicolumn{1}{c}{Wave 1} & \multicolumn{1}{c}{Wave 2} \\  
  \midrule
   Intercept & $3.553^{***}$ & $3.210^{***}$ & $-3.518^{***}$ & $-3.923^{***}$ \\ 
            & $(0.260)$      & $(0.245)$      & $(0.254)$   & $(0.263)$ \\ [0.5ex]
   $PE_{offline}$ & $-0.976^{***}$ & $-0.935^{***}$ & $0.707^{***}$ & $1.029^{***}$ \\ 
            & $(0.066)$            &    $(0.066)$   & $(0.062)$ & $(0.071)$ \\  [0.5ex]
   $PE_{online}$ & $-0.366^{***}$ & $-0.318^{***}$ & $0.493^{***}$ &  $0.266^{***}$\\ 
            & $(0.065)$           & $(0.062)$               & $(0.066)$ &  $(0.066)$\\ [0.5ex]
   Metro area & $-0.357^{**}$ & $-0.330^{*}$ & $-0.460^{***}$ &  $-0.381^{**}$\\ 
            & $(0.136)$       &   $(0.133)$  & $(0.132)$ &  $(0.134)$\\  [0.5ex]
   Ethnicity: White & $0.626^{***}$ & $0.722^{***}$ & $0.426^{**}$ & $0.446^{**}$ \\ 
            & $(0.147)$             & $(0.145)$ & $(0.142)$ & $(0.147)$ \\ [0.5ex]
   Ethnicity: Black & $-1.202^{***}$ & $-1.334^{***}$ & $-1.455^{***}$ & $-1.303^{***}$ \\ 
            & $(0.287)$              & $(0.293)$      & $(0.282)$ & $(0.285)$ \\ [0.5ex]
   Ethnicity: Hispanic & $0.453^{*}$ & $0.530^{**}$ & $0.535^{**}$ & $0.626^{**}$ \\ 
            & $(0.194)$              & $(0.192)$    & $(0.184)$ & $(0.192)$ \\ [0.5ex]
   Education & $-0.194$ & $-0.234^{*}$ & $-0.393^{***}$ & $-0.426^{***}$ \\ 
            & $(0.110)$ &  $(0.110)$   & $(0.106)$ & $(0.109)$ \\ [0.5ex]
  \midrule 
     $AIC$ & $2942.8$ & $2969.1$ & $3110.5$ & $2981.8$ \\
     N & $2420$ & $2420$ & $2420$ & $2420$ \\
  \bottomrule
  \multicolumn{5}{@{}l@{}}{\footnotesize Note: $^{*}\, p<.05$; $^{**}\, p<.01$; $^{***}\, p<.001$}
  \end{tabular}
  }
\end{table}

\clearpage
\bibliographystyle{plain}
\bibliography{sample}